\documentclass[11pt]{article}
\usepackage[T1]{fontenc}
\usepackage[utf8]{inputenc}
\usepackage{lmodern}
\usepackage{amsthm,amssymb,amsfonts,amsmath,amscd}
\usepackage{nccmath}
\usepackage{graphicx}
\usepackage{microtype}
\usepackage[pdfa,hidelinks]{hyperref}

\makeatletter \@addtoreset{equation}{section} \makeatother

\newtheorem{theorem}{Theorem}[section]
\newtheorem{lemma}{Lemma}[section]
\newtheorem{corollary}{Corollary}[section]

\newtheorem{proposition}{Proposition}[section]

\newcommand{\cF}{\mathcal{F}}
\newcommand{\cL}{\mathcal{L}}

\newcommand{\p}{\mathbb{P}}

\newcommand{\Tr}{\mathrm{Tr}\,}

\hypersetup{
 pdftitle={Eigenvector Occupation Statistics and Extremes in the Rosenzweig--Porter Model},
 pdfauthor={Yan V. Fyodorov, Mariya Shcherbina, Tatyana Shcherbina},
 pdfsubject={Eigenvector occupation statistics in the GUE Rosenzweig--Porter model},
 pdfkeywords={Rosenzweig--Porter model, eigenvectors, multifractality, Poisson point processes, extreme-value statistics}
}

\begin{document}

\title{Eigenvector Occupation Statistics and Extremes\\
in the Rosenzweig--Porter Model}
\author{Yan V. Fyodorov$^{1}$, Mariya Shcherbina$^{2,3}$, and Tatyana Shcherbina$^{4}$\\[0.7em]
\small $^{1}$Department of Mathematics, King's College London, London WC2R 2LS, United Kingdom\\
\small $^{2}$School of Mathematics, University of Bristol, Bristol BS8 1UG, United Kingdom,\\
 \small $^{3}$ B.~Verkin Institute for Low Temperature Physics and Engineering,\\[-0.2em]
\small National Academy of Sciences of Ukraine, Kharkiv 61103, Ukraine\\
\small $^{4}$ Department of Mathematics, University of Wisconsin--Madison, Madison, WI 53706, USA}
\date{}
\maketitle

\begin{abstract}
We study the occupation probabilities $p_i=|u_i|^2$ of a normalized bulk
eigenvector in the GUE Rosenzweig--Porter model
$A_N=B_N+\tau_NG_N$, with $\tau_N^2=N^{1-\gamma}$ and
$1<\gamma<2$.  Conditioning on an eigenvalue at a regular bulk energy $E$
through the normalized one-eigenpair Kac--Rice law, we derive an exact finite-$N$ projective density and reduce its
asymptotic analysis to a
Hubbard--Stratonovich integral over \(2\times2\) Hermitian matrices, which we evaluate by the saddle-point
method.  The
resulting occupation field is governed by a self-consistent Lorentzian
profile with effective resonant size $K_N(E)\asymp N^{2-\gamma}$ and local
exponential fluctuations.  We prove the asymptotics of all generalized
inverse participation sums and identify a transition at $q=1/2$.  We also
obtain a two-branch thickened amplitude-counting spectrum.  At the two ends
of the occupation field we prove marked Poisson point-process limits: the
largest components have Gumbel fluctuations, whereas the smallest component
has an exponential limit on the scale $N^{-1-\gamma}$.  Finally, for every
fixed $q>0$ the largest component contributes an asymptotically vanishing
fraction of the inverse participation sum, showing that these fixed-$q$
statistics do not freeze.  The statements are quenched, uniformly on a
high-probability regularity event for the diagonal disorder, and hence also
hold in the annealed setting.
\end{abstract}

\noindent\textbf{Keywords.} Rosenzweig--Porter model; eigenvectors;
occupation probabilities; inverse participation ratios; multifractality;
Poisson point processes; extreme-value statistics; Kac--Rice formula.

\medskip
\noindent\textbf{Mathematics Subject Classification (2020).}
60B20, 60G70, 82B44.

\tableofcontents

\section{Introduction}
\label{sec:introduction}

The Rosenzweig--Porter (RP) ensemble is one of the simplest
random-matrix models exhibiting non-ergodic delocalization.  To explain
this terminology, recall that a typical bulk eigenvector of an invariant
random-matrix ensemble is ergodic: its norm is spread over $O(N)$
coordinates, with typical occupation weights $p_i=O(N^{-1})$.  At the
opposite extreme, in the localized regime of a random Schr\"odinger
operator, an eigenfunction has appreciable weight on only $O(1)$ sites.
Critical wave functions at an Anderson transition provide an intermediate
possibility, in which the number of significantly occupied sites diverges
with the system size but remains sublinear in the total volume
\cite{WegnerIPR,MirlinEvers}.  The RP ensemble realizes an analogous
intermediate regime in a particularly tractable setting, without an
underlying spatial geometry.  In its non-ergodic delocalized phase, an
eigenvector occupies a diverging but vanishing fraction of the $N$
available coordinates.  The model therefore isolates the basic resonance
mechanism responsible for such states while remaining simple enough to
permit a detailed analysis of the full occupation statistics.

We consider
\[
 A_N=B_N+\tau_NG_N,\qquad
 B_N=\operatorname{diag}(b_1,\ldots,b_N),\qquad
 \tau_N^2=N^{1-\gamma},\qquad 1<\gamma<2,
\]
where $G_N$ is GUE and the $b_i$ are independent samples from a regular
density $\rho$.  If $u$ is a normalized eigenvector, its occupation weights
\[
 p_i=|u_i|^2,\qquad \sum_{i=1}^Np_i=1,
\]
form a random probability measure on the diagonal sites.  The generalized
inverse participation sums
\[
I_q=\sum_{i=1}^N p_i^q
\]
record moments of this measure.  In terminology customary in physics, the
distribution of components over the polynomial scales
$p_i\asymp N^{-\alpha}$ is encoded in the multifractal, or singularity,
spectrum $f(\alpha)$.  Mathematically, we formulate this spectrum through
``thickened'' amplitude counts: one counts the components in the logarithmic
window
\[
N^{-\alpha-\varepsilon}\leq p_i\leq N^{-\alpha+\varepsilon}
\]
and lets $\varepsilon\downarrow0$ after taking $N\to\infty$.

The purpose of this paper is to describe all three aspects at a fixed regular
bulk energy: the generalized inverse participation sums, the multifractal
spectrum $f(\alpha)$, and the upper and lower extreme-value statistics of the
occupation weights, including the largest and smallest weights
\[
p_{\max}=\max_i p_i,
\qquad
p_{\min}=\min_i p_i.
\]

\subsection{Background and motivation}

Rosenzweig and Porter introduced their ensemble as an interpolation between
Poisson and Wigner--Dyson level statistics in complex atomic spectra
\cite{RosenzweigPorter}.  The spectral crossover in the RP ensemble was subsequently studied by
several complementary random-matrix methods
\cite{AltlandJanssenShapiro,KunzShapiro}.  Closely related exact formulas
for spectral correlations in the external-source ensemble, consisting of
a deterministic Hermitian matrix perturbed by a GUE matrix, were derived
in \cite{BrezinHikami}.  Conditioned on its diagonal
part, the RP ensemble has precisely this deformed-GUE structure, although
with an $N$-dependent perturbation strength. In the generalized RP scaling, where the strength of the Gaussian
perturbation depends on $N$, the phase diagram comprises an ergodic
phase, a localized phase, and an intermediate non-ergodic delocalized
phase.  This three-phase structure was first identified systematically
by Kravtsov, Khaymovich, Cuevas et al. in \cite{KravtsovKhaymovichCuevasAmini}. In the convention of the present paper,
the intermediate phase is $1<\gamma<2$ and the characteristic number of
resonant coordinates is of order
\[
 N^{2-\gamma}.
\]
Thus a bulk eigenvector is extended over a diverging number of sites but over
a vanishing fraction of the ambient space.

The physics literature has developed a detailed description of RP
eigenfunctions using resolvent, perturbative, replica, supersymmetric, and
dynamical methods.  In particular, the Lorentzian resonant profile and the
participation exponents were obtained in
\cite{FacoettiVivoBiroli,TruongOssipov,Monthus}, while Bogomolny and Sieber
derived an explicit one-component eigenfunction distribution from the
Breit--Wigner profile and a local Porter--Thomas ansatz
\cite{BogomolnySieber}.  Replica and dynamical aspects were further studied in
\cite{DeTomasiSurvival,KhaymovichKravtsovDynamical,VenturelliCugliandoloSchehrTarzia},
and the local miniband viewpoint was developed in
\cite{AltshulerKravtsovCantor}.

It is useful to distinguish clearly between those of our results that
rigorously confirm this physical picture and those that provide information
beyond it. The scaling exponents of the inverse participation sums, including their
change of behaviour at $q=1/2$, and the conventional RP branch
\[
f(\alpha)=1-\frac{\gamma}{2}+\frac{\alpha}{2},
\qquad
2-\gamma\leq\alpha\leq\gamma,
\]
were already predicted in the physics literature.  Moreover, for the
Gaussian RP ensemble at the centre of the spectrum, Bogomolny and Sieber
obtained explicit leading prefactors and certain finite-$N$ corrections to
the disorder-averaged moments \cite{BogomolnySieber}.  We therefore do not
claim a more precise asymptotic expansion of the annealed inverse
participation sums.  Our result is different in nature: it establishes the
corresponding quenched statement.  For a typical fixed realization of the
diagonal disorder, the actual inverse participation sums concentrate around
their explicitly determined leading values, uniformly over regular bulk
energies.

A one-component distribution, however, does not determine the joint
rare-event statistics of all $N$ occupation weights.  In particular, although
a Gumbel law for the maximum may be suggested by a heuristic picture of
approximately independent tails, it does not follow from the one-component
distribution alone.  Correlations between the components, their
site-dependent scales, and the common normalization $\sum_i p_i=1$ may all
affect the extreme-value statistics.  Our results establish the Poisson
convergence of the upper extreme process and thereby prove the Gumbel law for
the maximum, including its precise centering.  They also establish the Poisson limit at the lower edge, the resulting law of
the minimum, and, jointly with the extreme weights, determine the distribution of the diagonal energies of the sites on which
these extreme weights occur. This global and extreme-value information is the principal
new content of the present paper.

Several rigorous results establish important parts of the deformed-Wigner and
RP picture.  Von Soosten and Warzel proved non-ergodic delocalization on the
expected resonant scale \cite{vSW}.  Benigni obtained asymptotic Gaussianity
and a deformed quantum unique ergodicity statement for mesoscopic deformed
Wigner matrices \cite{Benigni}.  Related perturbative spectral-measure results
were proved by Benaych-Georges, Enriquez, and Micha\"il \cite{BGEM}, while
M\'esz\'aros and Vir\'ag analysed Gaussian-wave limits for a concrete deformed
model \cite{MeszarosVirag}.  Very recent work of Cipolloni, Erd\H{o}s, and Henheik treats local  laws,
thermalization, and localization questions for a broad RP-type class
\cite{CEH}.  The present paper complements this literature by obtaining
occupation observables and extreme-value point processes directly from a
one-eigenpair law.

At the level of model building, the RP ensemble stands in relation to
multifractal random Schr\"odinger operators much as Derrida's Random
Energy Model (REM) stands in relation to logarithmically correlated
random-energy landscapes.  
The REM provides a solvable independent-energy backbone which retains the
competition between the entropy of the available states and the statistical
advantage of rare energies \cite{DerridaREM}. Logarithmic REMs enrich this picture by restoring correlations across
scales, thereby producing a more structured multifractal and
extreme-value phenomenology
\cite{FyodorovMultifractalityReview,FyodorovGiraud}.  The same
log-correlated freezing paradigm also appears in the extreme-value
statistics of random-matrix characteristic polynomials and has been
conjecturally extended to the Riemann zeta function
\cite{FyodorovKeating}.

In the same way, the RP ensemble provides a geometry-free, all-to-all
backbone for resonance-driven eigenfunction statistics.  It retains the
competition between the number of available sites and the enhancement
produced by near resonances, but suppresses the spatial organization and
correlations present in genuine random Schr\"odinger operators.  From
this perspective, spatially correlated extensions of the RP mechanism
play a role analogous to logarithmically correlated extensions of the
REM: they provide a bridge from the solvable mean-field model towards the
richer multifractality of spatially disordered systems. On the Schr\"odinger side, a recent rigorous result by Keating and
Uebersch\"ar proves, in dimensions $d\geq5$, an explicit nonconstant
spectrum of R\'enyi exponents for an averaged Fourier-space measure
associated with eigenfunctions of Schr\"odinger operators with point-like
scatterers in the thermodynamic limit, under suitable weak-disorder and
strong-coupling assumptions \cite{KeatingUeberschaer}.  Complementary
routes towards richer, genuinely multifractal regimes have been
investigated in RP-inspired ensembles with structured disorder and
correlated or spatially decaying hopping
\cite{KutlinKhaymovich,RoyBasuKhaymovich}.

\subsection{Main results and method}

Our starting point is the normalized Kac--Rice Palm law of an eigenvector
conditioned on the occurrence of an eigenvalue at energy $E$.  For a fixed
diagonal realization $B_N$, we derive an exact finite-$N$ density on complex
projective space.  A fermionic representation of the tangent determinant and
a Hubbard--Stratonovich transform reduce its asymptotic analysis to an integral over $2\times2$ Hermitian matrices.
 The idea of using the Kac--Rice counting formula to derive joint
eigenvalue--eigenvector densities for random matrices was developed in
\cite{Fyo:KR}, initially in the non-Hermitian setting.  Here we adapt this
construction to the Hermitian Rosenzweig--Porter ensemble and to the
projective law of an eigenvector conditioned on an eigenvalue at the fixed
energy $E$.  The subsequent fermionic and Hubbard--Stratonovich reduction is
closely related to the supersymmetric method \cite{Ef};
the mixed-sign saddle analysis is also closely related to characteristic-
polynomial asymptotics in \cite{ShcherbinaWignerCharPoly,ShcherbinaMixedMoment1D,MSTLocalizationSide}.

The self-consistent saddle produces deterministic weights
\[
 m_i(E)=\frac{\tau_N^2}
 {N\bigl((b_i-E+t_N(E))^2+\eta_N(E)^2\bigr)},
 \qquad \sum_i m_i(E)=1,
\]
with $\eta_N(E)\sim\pi\rho(E)\tau_N^2$.  Their effective resonant number is
\[
 K_N(E)=\frac{N\eta_N(E)^2}{\tau_N^2}
 \sim \pi^2\rho(E)^2N^{2-\gamma}.
\]
At the observable level, the occupation weights behave as the Lorentzian
profile multiplied by local mean-one exponential variables.  The exact
normalization constraint is retained throughout the Kac--Rice calculation.

The consequences are as follows.

\begin{enumerate}
\item For every fixed $q>0$, the generalized inverse participation sum
$I_q=\sum_i p_i^q$ is asymptotic to
$\Gamma(q+1)\sum_i m_i(E)^q$.  The deterministic sum changes mechanism at
$q=1/2$: resonant coordinates dominate for $q>1/2$, while the bulk of
off-resonant coordinates contributes for $q<1/2$; the critical case has a
logarithmic correction.

\item The number of coordinates with $p_i\asymp N^{-\alpha}$ has a
two-branch exponent $f(\alpha)$.  The increasing branch comes from the
Lorentzian spatial profile, whereas the decreasing branch is generated by
the small-value tail of the local exponential factor.  This gives a
degenerate piecewise-linear form of multifractality, rather than merely a
single support dimension.

\item The upper and lower extremes have different mechanisms.  After
rescaling, the largest occupations form a marked Poisson process and their
maximum has a Gumbel limit, with a $-\frac12\log\log K_N$ correction to the
centering.  The smallest occupations form a second marked Poisson process and
$N^{1+\gamma}\mu_2(E)p_{\min}$ converges to a mean-one exponential law.  The
associated diagonal marks distinguish the resonant origin of the upper
extremes from the off-resonant origin of the lower ones.
\end{enumerate}

Despite the nontrivial upper extremes, there is no condensation at fixed
$q$: we prove $p_{\max}^q/I_q\to0$ in probability for every fixed $q>0$.
Thus the Gumbel maximum does not produce logREM-type freezing of the physical
occupation vector.

All asymptotic statements are first proved quenched, uniformly on a
high-probability regularity event for $B_N$, and therefore pass to the
annealed law.  Section~\ref{s:1} defines the model, the Kac--Rice Palm law, and
states the results.  Section~\ref{sec:of-exact} derives the exact one-frame
density and its integral representation.  Section~\ref{s:2} performs the
saddle analysis and proves the limit theorems.  The Appendix collects the
uniform empirical and self-consistency estimates used in the proof.

\section{Model}\label{s:1}
Let $$A_N=B_N+\tau_NG_N,$$ where $B_N=\operatorname{diag}(b_1,\ldots,b_N)$
and $G_N$ is a GUE matrix.
Assume 
\begin{equation}\label{tau}
\tau_N^2=N^{1-\gamma}, \quad 1<\gamma<2.
\end{equation}

The diagonal variables $b_1,\ldots,b_N$ are iid with density
$\rho$.  We assume that $\rho\in C^3(\mathbb R)$ and decays sufficiently fast at infinity to have a well-defined Hilbert transform,
and $\rho(x)>0$ on a fixed open interval containing the compact bulk interval $I$ under consideration.  
We also assume that  $\rho$ has a finite second moment and denote
\begin{equation}\label{m_2}
\mu_2(E)=\int (b-E)^2\rho(b) db.
\end{equation}


For $E\in I$, let $t_N(E)$ and $\eta_N(E)>0$ be the solution of
\begin{equation}\label{of:pastur}
 t_N=\frac{\tau_N^2}{N}\sum_{i=1}^N
 \frac{E-t_N-b_i}{(b_i-E+t_N)^2+\eta_N^2},
 \qquad
 1=\frac{\tau_N^2}{N}\sum_{i=1}^N
 \frac1{(b_i-E+t_N)^2+\eta_N^2}.
\end{equation}
As we will prove below, such a solution exists and is unique in the region $
|t|\le C\tau_N^2$, $c\tau_N^2\le\eta\le C\tau_N^2$, which is the regime used below.

Put
\begin{equation}\label{of:weights}
 m_i(E)=\frac{\tau_N^2}
 {N[(b_i-E+t_N(E))^2+\eta_N(E)^2]},
 \qquad \sum_{i=1}^Nm_i(E)=1,
 \qquad K_N(E)=\frac{N\eta_N(E)^2}{\tau_N^2}.
\end{equation}
We will also prove
\begin{align}\label{K_N}
 \eta_N(E)=\pi\rho(E)\tau_N^2[1+o(1)],\qquad
 K_N(E)=\pi^2\rho(E)^2N\tau_N^2[1+o(1)]\longrightarrow\infty.
\end{align}

\subsection{Main results}
Let $B_N=\text{diag}(b_1, \ldots, b_N)$ be fixed. Let also $d\omega_{FS}$ be the normalized Fubini--Study measure
on $\mathbb CP^{N-1}=\{u\in\mathbb C^N:\|u\|=1\}/\{u\sim e^{i\theta}u\}
$: it is the push-forward of normalized Haar measure
$d\sigma$ on the unit sphere $S^{2N-1}\subset\mathbb C^N$ under
$u\mapsto[u]$, where $[u]$ is the eigenvector line with the irrelevant complex
phase quotiented out.
Define the one-eigenpair intensity by
\[
 \mathbb E\sum_{\alpha=1}^NF(\lambda_\alpha,[\psi_\alpha])=
 \int F(E,[u])\rho^{(1)}_{B_N}(E,[u])\,dE\,d\omega_{FS}([u]),
\]
and the Kac-Rice law
\begin{equation}\label{KR_law}
d\mu_{N,E}^{KR}([u])=\dfrac{\rho^{(1)}_{B_N}(E,[u])d\omega_{FS}([u])}{\int_{\mathbb CP^{N-1}}\rho^{(1)}_{B_N}(E,[u])d\omega_{FS}([u])}.
\end{equation}
Let $u$ be sampled from this law and
\begin{equation}\label{p}
p_i=|u_i|^2.
\end{equation}

Let $\Omega_N^{\mathrm{loc}}$ be the event in the probability space of
$B_N$ on which Proposition \ref{p:LLN} and Lemma \ref{l:sol} hold.  Its
complement has probability at most $N^{-D}$ for every fixed $D>0$.
Choose a deterministic sequence $\delta_N\downarrow0$ such that
\begin{align}\label{mu_2.1}
& \mathbb P\left\{\sup_{E\in I}\left|
 \frac1N\sum_{i=1}^N(b_i-E)^2-\mu_2(E)\right|>\delta_N\right\}\longrightarrow0,\\
& \mathbb P\left\{\sup_{E\in I}\sup_{\|F\|_\infty+\operatorname{Lip}(F)\leq1}
 \left|\frac1N\sum_{i=1}^N(b_i-E)^2F(b_i)
 -\int(b-E)^2F(b)\rho(b)\,db\right|>\delta_N\right\}\longrightarrow0,
\notag\end{align}
and, after making $\delta_N$ decrease more slowly if necessary, also
\begin{align}\label{max_b}
\mathbb P\left\{\frac1N\max_{1\leq i\leq N}b_i^2>\delta_N\right\}
 \longrightarrow0.
\end{align}
Let $\Omega_N^{(2)}$ be the intersection of the three corresponding good events.
Such a choice is possible by the strong law of large numbers, compactness of
$I$, a standard truncation argument for the weighted empirical measures, and
the consequence of $\mathbb E b_1^2<\infty$ that
$N^{-1}\max_i b_i^2\to0$ in probability.  Finally, set
\[
 \Omega_N:=\Omega_N^{\mathrm{loc}}\cap\Omega_N^{(2)}.
\]
Thus $\mathbb P(\Omega_N^c)\to0$.

Then we have the following theorems
\begin{theorem}[Inverse participation sums]
\label{t:IPR}
Fix $q>0$ and set
\begin{align}\label{I_q}
I_q=\sum p_j^q
\end{align}
On the event $\Omega_N$, under the quenched
Kac--Rice  law, we have
\begin{equation}\label{IPR}
 \frac{I_q}{\Gamma(q+1)\sum_{i=1}^Nm_i(E)^q}\longrightarrow1
 \qquad\hbox{in probability}.
\end{equation}
The convergence is uniform in $B_N\in\Omega_N$ and therefore also holds under
the annealed law. 
The sum in the denominator has the following asymptotics:
\begin{align}
 \sum_i m_i^q
 &=K_N^{1-q}\frac{\Gamma(q-1/2)}{\sqrt\pi\,\Gamma(q)}[1+o_{\p}(1)],
 &&q>\frac12,\label{IPR-supercritical}\\
 \sum_i m_i^{1/2}
 &=\frac2\pi K_N^{1/2}\log\frac1{\eta_N}[1+o_{\p}(1)],
 &&q=\frac12,\label{IPR-critical}\\
 \sum_i m_i^q
 &=N^{1-\gamma q}\left[\int_{\mathbb R}|x-E|^{-2q}\rho(x)\,dx+o_{\p}(1)\right],
 &&0<q<\frac12.\label{IPR-subcritical}
\end{align}
All asymptotic relations here are in probability.
\end{theorem}

For $\alpha\in\mathbb R$ and $\varepsilon>0$, let
\[
 \mathcal N_N(\alpha,\varepsilon)=\#\{i:N^{-\alpha-\varepsilon}
 \leq p_i\leq N^{-\alpha+\varepsilon}\}.
\]

\begin{theorem}[Thickened amplitude-counting spectrum]
\label{thm:ampl}
On $\Omega_N$, for every $\alpha$ in the interior of a region where the function below is
positive, under the quenched
Kac--Rice  law, we have
\[
 \lim_{\varepsilon\downarrow0}\lim_{N\to\infty}
 \frac{\log\mathcal N_N(\alpha,\varepsilon)}{\log N}=f(\alpha)
 \quad\hbox{in probability},
\]
where
\[
 f(\alpha)=
 \begin{cases}
 1-\gamma/2+\alpha/2,&2-\gamma\leq\alpha\leq\gamma,\\
 1+\gamma-\alpha,&\gamma<\alpha\leq\gamma+1.
 \end{cases}
\]
The convergence is uniform in $B_N\in\Omega_N$ and therefore also holds under
the annealed law. 
\end{theorem}

\paragraph{Physical origin of the two branches.}
The two branches in Theorem~\ref{thm:ampl} arise from two different
mechanisms.  For
\[
2-\gamma\leq\alpha\leq\gamma,
\]
the scale $m_i(E)\asymp N^{-\alpha}$ is produced by diagonal energies at
distance
\[
|b_i-E|\asymp N^{-(\gamma-\alpha)/2}
\]
from the observation energy.  The expected number of such sites is therefore
of order
\[
N\,N^{-(\gamma-\alpha)/2}
=
N^{\,1-\gamma/2+\alpha/2}.
\]
This gives the left, blue branch
\[
f_{\mathrm L}(\alpha)
=
1-\frac{\gamma}{2}+\frac{\alpha}{2},
\]
which is the conventional RP multifractal spectrum already discussed in the
physics literature.

The right, orange branch has a different origin.  At a generic off-resonant
site one has $m_i(E)\asymp N^{-\gamma}$, while the local GUE
Porter--Thomas factor is exponential:
\[
p_i\simeq m_i(E)\xi_i,
\]
where $\xi_i\sim{\rm Exp}(1)$, i.e.\ $\xi_i$ has the probability density $\mathcal P(\xi)\,d\xi=e^{-\xi}\,d\xi,
\qquad \xi\geq 0$.

To obtain a smaller occupation $p_i\asymp N^{-\alpha}$ with
$\alpha>\gamma$, one needs
\[
\xi_i\asymp N^{-(\alpha-\gamma)}.
\]
Since the exponential density is nonzero at the origin, among $N$ sites the
number of such small factors is of order
\[
N\,N^{-(\alpha-\gamma)}
=
N^{\,1+\gamma-\alpha}.
\]
This gives
\[
f_{\mathrm R}(\alpha)
=
1+\gamma-\alpha,
\qquad
\gamma<\alpha\leq\gamma+1.
\]
The endpoint $\alpha=\gamma+1$ is where the expected number of such
components becomes of order one.

\begin{figure}[t]
\centering
\includegraphics[width=0.82\textwidth]{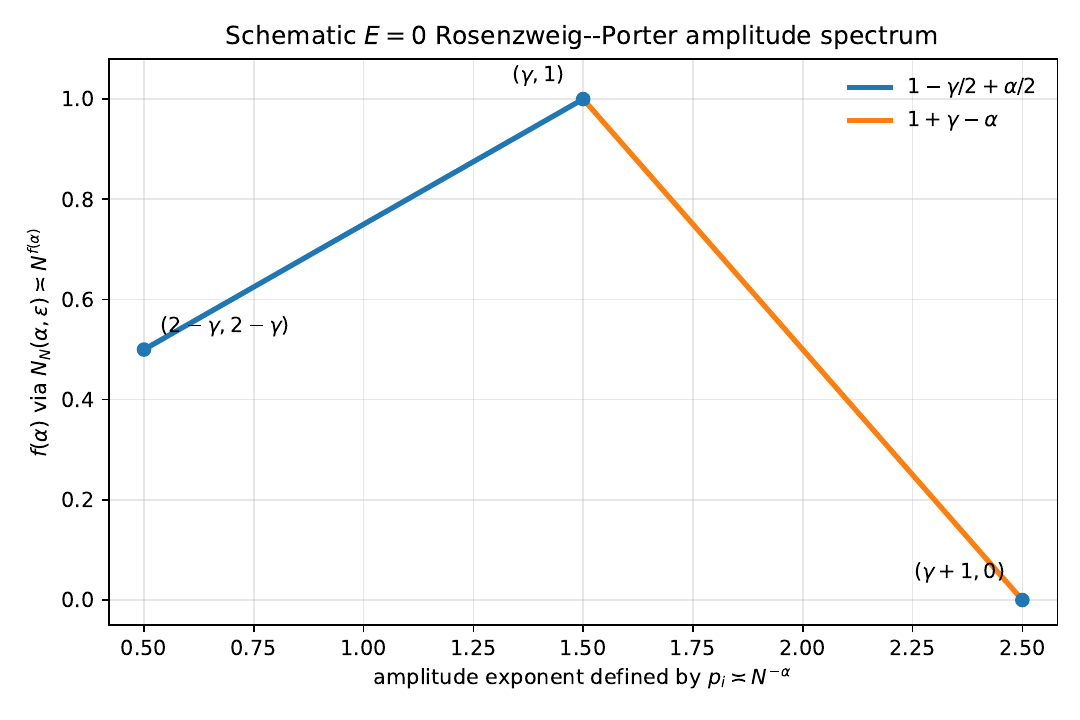}
\caption{The thickened amplitude-counting spectrum $f(\alpha)$ for
$\gamma=3/2$, drawn at $E=0$.  The left branch is the Lorentzian structural
spectrum, and the right branch is the small-amplitude exponential tail.  The
same exponents hold at every regular bulk energy covered by
Theorem~\ref{thm:ampl}; the energy dependence enters only through the
non-universal prefactors.}
\label{fig:falpha}
\end{figure}

The small-Porter--Thomas mechanism is implicit in the one-component
distribution of Bogomolny and Sieber \cite{BogomolnySieber}.  To the best of
our knowledge, however, the resulting right branch was not previously
isolated as part of the RP amplitude-counting spectrum.  The new statement
here is its explicit identification and theorem-level proof for the GUE
model, together with the subsequent Poisson limits for the upper and lower
extreme processes.  Figure~\ref{fig:falpha} displays the two branches for
$\gamma=3/2$.

\medskip

Let $\alpha_N$ be the unique solution, for large $N$, of
\begin{equation}\label{a_n_eq}
 \frac{K_N(E)e^{-\alpha_N}}{\sqrt{\pi \alpha_N}}=1.
\end{equation}
Equivalently,
\begin{equation}\label{a_N}
 \alpha_N=\log K_N(E)-\frac12\log\log K_N(E)-\frac12\log\pi+o(1).
\end{equation}
\begin{theorem}\label{t:max}
Define
\begin{equation}\label{beta,v}
\beta_i=b_i-E+t_N,\qquad
v_i=\sqrt{\alpha_N}\,\frac{\beta_i}{\eta_N},
\end{equation}
and the point process
\[
 \Xi_N^+
=\sum_{i=1}^N
 \delta_{(v_i,\,K_Np_i-\alpha_N)}.
\]
Then, under the quenched Kac--Rice law, uniformly in $B_N\in\Omega_N$ and $E\in I$,
\[
\Xi_N^+
 \Longrightarrow
 \operatorname{PPP}\!\left(
 \frac{e^{-v^2}}{\sqrt{\pi}}\,dv\,e^{-x}\,dx
 \right),
\]
vaguely on \(\mathbb R\times(-A,\infty)\), for every fixed $A$.
The same conclusion holds under the annealed law.
\end{theorem}
\begin{corollary}\label{c:max}
Let 
\[
p_{\text{max}}=\max\limits_{1\le i\le N} p_i.
\]
Then on $\Omega_N$, under the same quenched and annealed laws,
\[
p_{\text{max}}=\dfrac{1}{K_N(E)} (\log K_N(E)-\frac12\log\log K_N(E)-\frac12\log\pi+O_{\mathbb P}(1)).
\]
More precisely,
\[
K_Np_{\max}-\alpha_N \Longrightarrow \mathcal{G},
\]
where $\mathcal{G}$ has the Gumbel distribution
\[
\mathbb P\{\mathcal{G}\leq x\}=e^{-e^{-x}}.
\]
The convergence is uniform in $B_N\in\Omega_N$ and therefore also holds under
the annealed law. 
\end{corollary}

\begin{corollary}[Absence of freezing at fixed $q$]
\label{c:no-freezing}
For every fixed $q>0$, under the same quenched and annealed laws,
\begin{equation}\label{no-freezing}
 \frac{p_{\max}^q}{I_q}\xrightarrow{\mathbb P}0.
\end{equation}
Thus, although $p_{\max}$ has nontrivial Gumbel fluctuations on its natural
extreme-value scale, the largest component carries an asymptotically vanishing
fraction of every fixed positive inverse participation sum $I_q$.  In this
sense the fixed-$q$ statistics remain collective and do not freeze onto the
largest component.
\end{corollary}

\begin{proof}
Corollary \ref{c:max} gives
\[
 p_{\max}=O_{\mathbb P}\left(\frac{\log K_N}{K_N}\right).
\]
If $q>1/2$, Theorem \ref{t:IPR} yields
\[
 I_q=C_qK_N^{1-q}[1+o_{\mathbb P}(1)],
 \qquad
 C_q=\frac{\Gamma(q+1)\Gamma(q-1/2)}
 {\sqrt\pi\,\Gamma(q)}>0,
\]
and hence
\[
 \frac{p_{\max}^q}{I_q}
 =O_{\mathbb P}\left(\frac{(\log K_N)^q}{K_N}\right)=o_{\mathbb P}(1).
\]
For $q=1/2$, the critical asymptotics in Theorem \ref{t:IPR} give
\[
 \frac{p_{\max}^{1/2}}{I_{1/2}}
 =O_{\mathbb P}\left(
 \frac{(\log K_N)^{1/2}}
 {K_N\log(1/\eta_N)}\right)=o_{\mathbb P}(1).
\]
Finally, if $0<q<1/2$, then $K_N\asymp N^{2-\gamma}$ and
\[
 I_q=C_q(E)N^{1-\gamma q}[1+o_{\mathbb P}(1)],
 \qquad
 C_q(E)=\Gamma(q+1)\int_{\mathbb R}|x-E|^{-2q}\rho(x)\,dx>0.
\]
Therefore
\[
 \frac{p_{\max}^q}{I_q}
 =O_{\mathbb P}\left(
 (\log N)^qN^{-1+2q(\gamma-1)}\right)=o_{\mathbb P}(1),
\]
because $2q(\gamma-1)<1$.
\end{proof}

\medskip

\begin{theorem}\label{t:min}
Define
\begin{equation}\label{A_N}
\mathcal A_N(E):=c_NN\mu_2(E),
\end{equation}
with $\mu_2(E)$ defined in (\ref{m_2}),
and the point process
\[
 \mathcal L_N^-:=\sum_{i=1}^N\delta_{(\mathcal A_N(E)p_i,b_i)}.
\]
Then, under the quenched Kac--Rice law, uniformly in $B_N\in\Omega_N$ and $E\in I$,
\[
 \mathcal L_N^-\Longrightarrow
 \operatorname{PPP}\bigl(dx\,\nu_E(db)\bigr)
 \quad\hbox{vaguely on }(0,\infty)\times\mathbb R,
\]
where
\begin{equation}\label{lower-mark-law}
 \nu_E(db)=\frac{(b-E)^2\rho(b)}{\mu_2(E)}\,db.
\end{equation}
The same conclusion holds under the annealed law.
\end{theorem}

\begin{corollary}
\label{cor:min}
Let $p_{\min}=\min_i p_i$.  Under the same quenched and annealed laws,
\[
\mathcal A_N(E)p_{\min}\Longrightarrow {\rm Exp}(1),\]
or, equivalently,
 \[
 N^{1+\gamma}\mu_2(E)p_{\min}\Longrightarrow {\rm Exp}(1).
\]
\end{corollary}

%

\section{Exact one-frame Kac--Rice density}\label{sec:of-exact}
\subsection{Kac-Rice formula}
Let $u$ be the eigenvector corresponding to the eigenvalue $E$.
Set $P_u=I-u\otimes u^*$ and let $V_u:\mathbb C^{N-1}\to u^\perp$ be an
isometry.  Then for the Kac-Rice law \eqref{KR_law} we have
\begin{proposition}\label{p:of-KR}
If $K$ is an independent $(N-1)\times(N-1)$ GUE matrix with $d\p(K)\propto e^{-N\Tr K^2/2} dK$, then
\begin{align}\label{of:KR}
 \rho^{(1)}_{B_N}(E,[u])={}&C_{N,\tau_N}
 \exp\left\{-\frac{N}{\tau_N^2}\|P_uB_Nu\|^2
 -\frac{N}{2\tau_N^2}(E-u^*B_Nu)^2\right\}\nonumber\\
 &\times\mathbb E_K\left|\det\bigl(\tau_NK+V_u^*(B_N-E)V_u\bigr)\right|^2.
\end{align}
Moreover, denote $p_i=|u_i|^2$, $\tilde b_i=b_i-E$, and   
$$\Delta_{N-1}(p)=\{p_i\ge 0,  \,i=1,..,N, \sum\limits_{i=1}^N p_i=1\}.$$
Fixing $E$, (\ref{of:KR}) can be rewritten as
\begin{equation}\label{KR}
 d\mu_{N,E}^{KR}(p)\propto {\bf1}_{\Delta_{N-1}}(p)
 e^{-\tfrac{N}{\tau_N^2} Q_E(p)}D_{N,E}(p)\,dp,
\end{equation}
where $dp$ is Lebesgue measure,and
\begin{equation}\label{Q}
 Q_E(p)=\sum_i \tilde b_i^2p_i-\frac12\left(\sum_i \tilde b_ip_i\right)^2,
 \qquad
 D_{N,E}(p)=\mathbb E_K\left|\det(\tau_NK+V_u^*(B_N-E)V_u)\right|^2.
\end{equation}
\end{proposition}

\begin{proof} Fix $u$ with $u^*u=1$, and choose $V_u$ as above. Set $U=(u,V_u)\in U(N)$.
The key ingredient of this proposition is the projective Kac--Rice
eigenpair formula, which is the Hermitian specialization of the
eigenvalue--eigenvector density construction developed in
\cite{Fyo:KR} for several classes of non-Hermitian random matrices.
Up to a normalization constant depending only on the
chosen conventions for $d\omega_{FS}$ and the delta functions, the formula implies
\begin{equation*}
 \rho^{(1)}_{B_N}(E,[u])\propto\mathbb{E}_{G_N}\left[
 \delta_{\mathbb{R}}\bigl(u^*(A_N-E)u\bigr)
 \delta_{\mathbb{C}^{N-1}}\bigl(V_u^*(A_N-E)u\bigr)
 \left|\det\bigl(V_u^*(A_N-E)V_u\bigr)\right|^2\right].
\end{equation*}
By GUE invariance,
\[
 U^*G_NU=\begin{pmatrix}h&\xi^*\\ \xi&K\end{pmatrix},
\]
where $h\in\mathbb R$, $\xi\in\mathbb{C}^{N-1}$, and $K\in\text{Herm}(N-1)$ are independent, and
the scalar $h$ has density $e^{-Nh^2/2}$,  the complex vector $\xi$
has density $e^{-N\|\xi\|^2}$, and $K$ is
distributed as an $(N-1)\times(N-1)$ GUE matrix with $d\p(K)\propto e^{-N\Tr K^2/2} dK$.

In the same basis,
\[
 U^*(A_N-E)U=
 \begin{pmatrix}
 \tau_Nh+u^*Bu-E&\tau_N\xi^*+u^*BV_u\\
 \tau_N\xi+V_u^*Bu&\tau_N K+V_u^*(B-E)V_u
 \end{pmatrix}.
\]
  The eigenpair equations from the delta-functions  force
  \begin{align*}
 & \tau_Nh+u^*B_Nu-E=0\Longrightarrow h=\tau_N^{-1} (E-u^*B_Nu),\\
  & \tau_N\xi+V_u^*B_Nu=0\Longrightarrow \xi= -\tau_N^{-1}  V_u^*B_Nu.
  \end{align*}
Since $VV^*=P_u$, the vector term can be written as
\[
 \|V^*Bu\|^2=\|P_uBu\|^2.
\]  
This gives the
exponential factor in \eqref{of:KR}; the free quotient block has the
displayed determinant-square tangent Jacobian.  

To prove (\ref{KR}), write
\[
 u^*B_Nu=\sum_i b_ip_i=E+\sum_i \tilde b_ip_i,
 \qquad
 \|P_uB_Nu\|^2=\sum_i b_i^2p_i-
 \left(\sum_i b_ip_i\right)^2.
\]
Hence, we obtain
\begin{align*}
 \|P_uB_Nu\|^2+\frac12(E-u^*B_Nu)^2
 &=\sum_i \tilde b_i^2p_i-\frac12\left(\sum_i \tilde b_ip_i\right)^2
 =Q_E(p).
\end{align*}
Finally, changing the phases of $u$ conjugates the compression
$V_u^*(B_N-E)V_u$ by a unitary operator.  The GUE block $K$ is unitarily
invariant, so the determinant expectation depends only on $p$; denote it by
$D_{N,E}(p)$.  Substitution in \eqref{of:KR} gives \eqref{KR}.
\end{proof}

\subsection{Integral representation for
\texorpdfstring{$D_{N,E}(p)$}{D(N,E)(p)}}\label{sec:of-HS}

\begin{proposition}
\label{prop:of-D-HS}
The determinant factor $D_{N,E}(p)$ from (\ref{Q}) can be represented as
\begin{equation}\label{of:HS}
 D_{N,E}(p)=\mathcal C_{N,\tau_N}
\int_{\operatorname{Herm}(2)}e^{-\frac{N}{2\tau_N^2}\operatorname{Tr}Y^2}
 \prod_i\det(\tilde b_iI_2+iY)
 \det\left[\sum_ip_i(\tilde b_iI_2+iY)^{-1}\right]dY.
\end{equation}
\end{proposition}

\begin{proof}
Let $d=N-1$, and put
\[
 C=V_u^*(B_N-E)V_u.
\]
For two Grassmann fields $\Psi_\alpha=\{\psi_{\ell,\alpha}\}_l,\bar \Psi_\alpha=\{\bar\psi_{\ell,\alpha}\}_l$
($1\leq\ell\leq d$, $1\leq\alpha\leq2$), write the standard determinant representation via Grassmann integral
\[
 \int\exp\left\{\Psi^+ M \Psi\right\}
 \prod\limits_j d\bar\psi_j\,d\psi_j=\det M
\]
valid for any $d\times d$ matrix $M$.

It gives
\[
 |\det(\tau_NK+C)|^2
 =\int\exp\left\{\sum_{\alpha=1}^2
 \Psi_\alpha^+(\tau_NK+C)\Psi_\alpha\right\}\,d \Psi,\quad d\Psi= \prod\limits_{\alpha=1}^2\prod\limits_{j=1}^d d\bar\psi_{j,\alpha}\,d\psi_{j,\alpha}.
\]
Define the Grassmann-even $d\times d$ matrix
\[
 R_{jk}:=\sum_{\alpha=1}^2\bar\psi_{j,\alpha}\psi_{k,\alpha},
\]
and the $2\times2$ matrix
\[
 S_{\alpha\beta}:=\sum_{j=1}^d
 \bar\psi_{j,\alpha}\psi_{j,\beta}.
\]
The GUE average gives
\[
 \mathbb E_K e^{\tau_N\operatorname{Tr}(KR)}
 =\exp\left\{\frac{\tau_N^2}{2N}\operatorname{Tr}R^2\right\}.
\]
Anticommutation of the Grassmann variables yields
\[
 \operatorname{Tr}R^2=-\operatorname{Tr}S^2.
\]
Consequently,
\[
 D_{N,E}(p)
 =\int\exp\left\{\sum_{\alpha=1}^2\Psi_\alpha^+ C\Psi_\alpha
 -\frac{\tau_N^2}{2N}\operatorname{Tr}S^2\right\}\,d\Psi.
\]
Applying the Hubbard-Stratonovich transformation (see, e.g., \cite{Ef}), we get
\[
 e^{-\frac{\tau_N^2}{2N}\operatorname{Tr}S^2}
 =C_{N,\tau_N}\int_{\operatorname{Herm}(2)}
 e^{-\frac{N}{2\tau_N^2}\operatorname{Tr}Y^2+i\operatorname{Tr}(YS)}\,dY.
\]
Now we can integrate over the Grassmann variables and obtain
\begin{equation}\label{of:uncompressed-HS}
 D_{N,E}(p)=C_{N,\tau_N}\int_{\operatorname{Herm}(2)}
 e^{-\frac{N}{2\tau_N^2}\operatorname{Tr}Y^2}
 \det\bigl(C\otimes I_2+iI_d\otimes Y\bigr)\,dY.
\end{equation}
It remains to evaluate the determinant.  Put
\[
 A_E(Y)=(B_N-E)\otimes I_2+iI_N\otimes Y,
 \qquad Q_u z=u\otimes z,\quad z\in\mathbb C^2.
\]
The range of $V_u\otimes I_2$ is the orthogonal complement of
$Q_u\mathbb C^2$.  The finite-dimensional compression determinant identity
\[
 \det\bigl((V_u\otimes I_2)^*A_E(Y)(V_u\otimes I_2)\bigr)
 =\det A_E(Y)\,\det\bigl(Q_u^*A_E(Y)^{-1}Q_u\bigr)
\]
therefore applies whenever $A_E(Y)$ is invertible.  Since $A_E(Y)$ is
diagonal in the site index,
\[
 \det A_E(Y)=\prod_i\det(\tilde b_i I_2+iY),
 \qquad
 Q_u^*A_E(Y)^{-1}Q_u
 =\sum_i p_i(\tilde b_iI_2+iY)^{-1}.
\]
Substituting this into \eqref{of:uncompressed-HS} gives \eqref{of:HS}.  
\end{proof}

\subsection{Final representation}
To simplify formulas below we set
\begin{align}\label{c_N}
c_N=N\tau_N^{-2}.
\end{align}
The positive quadratic term in \eqref{KR}
can be linearized as
\begin{equation}\label{r-HS}
 e^{\frac{c_N}{2}(\sum_i \tilde b_ip_i)^2}
 =\left(\frac{c_N}{2\pi}\right)^{1/2}
 \int_{\mathbb R}e^{-c_Nr^2/2+c_Nr\sum_i \tilde b_ip_i}\,dr.
\end{equation}
Together with the remaining term $e^{-c_N\sum_i \tilde b_i^2p_i}$ it gives 
$$e^{-c_N\sum_i (\tilde b_i^2-r\tilde b_i)p_i-c_Nr^2/2}.$$
Now to replace the simplex by positive independent variables, we use the following representation of the $\delta$-function
\begin{equation}\label{Bromwich}
 \delta\left(\sum_ip_i-1\right)=\frac{c_N}{2\pi i}
 \int_{\Gamma}e^{c_N\zeta(1-\sum_ip_i)}d\zeta
\end{equation}
with a vertical line $\Gamma$ from $\zeta_*-i\infty$ to $\zeta_*+i\infty$ which we will choose later.
Then the KR measure can be rewritten as
\begin{align*}
d\mu_{N,E}^{KR}(p)\propto\left(\int_{\operatorname{Herm}(2)}\int_{\mathbb R} \int_{\Gamma}  \right.&e^{-c_N(r^2/4-\zeta+\operatorname{Tr}Y^2/2)-c_N\sum_i p_i \Big((\tilde b_i-r/2)^2+\zeta\Big)}\\
&\left.\times\prod_i\det(\tilde b_iI_2+iY)
 \det\left[\sum_ip_i(\tilde b_iI_2+iY)^{-1}\right] dY dr d\zeta\right) dp,
\end{align*}
where we have used that
\[
\tilde b_i^2-r\tilde b_i=(\tilde b_i-r/2)^2-r^2/4,
\]
and that
\[
\sum\limits_ir^2p_i/4=r^2/4
\]
on a simplex $\Delta_{N-1}$.

Changing $Y=W^*\hat Y W$ with $2\times 2$ unitary $W$, and $\hat Y=\text{diag}\{y_1,y_2\}$  with real $y_1$, $y_2$, and using $dY\propto (y_1-y_2)^2 dy_1dy_2 dW$,
we get further after integrating over $dW$
\begin{multline}\label{mu^KR}
d\mu_{N,E}^{KR}(p)\propto\left(\int_{\mathbb R^2}\int_{\mathbb R} \int_{\Gamma}  e^{c_N(\cL_N(y_1)+\cL_N(y_2)+\Phi_N(\bar p,\zeta,r))}\right.\\
 \times \left. (y_1-y_2)^2\prod\limits_{\alpha=1}^2h(\bar p,y_\alpha) dy_1 dy_2 dr d\zeta\right) dp,
\end{multline}
with
\begin{align}\label{Phi,L,h}
&\Phi_N(\bar p,\zeta,r)= -\sum p_j((\tilde b_j-r/2)^2+\zeta)+\zeta-r^2/4\\
&\cL_N(y)=c_N^{-1}\sum\limits_j\log(\tilde b_j+iy)-y^2/2,\notag \\ 
& h(\bar p,y)=\sum_jp_j(\tilde b_j+iy)^{-1}.\notag 
\end{align}
Let
\[ 
a_i(\zeta,r)= (\tilde b_i-r/2)^2+\zeta.
\]
If we factor $e^{-c_N\sum_ia_i(\zeta,r) p_i} dp$ in $p_i$ to get independent exponential measures, we get the scalar action
\begin{equation}\label{F}
 \cF_N(r,\zeta,\bar y)=-\frac{r^2}{4}+\zeta
 -c_N^{-1}\sum_{i=1}^N\log a_i(r,\zeta)
\end{equation}
multiplied by $c_N$.

  More precisely, if
\[
 P_i^{(r,\zeta)}\sim\operatorname{Exp}(c_Na_i(r,\zeta))
\]
are independent, the positive-orthant integral is factored exactly by
\[
 \int_{\mathbb R_+^N}e^{-c_N\sum_i a_i(r,\zeta)p_i}F(p)\,dp
 =\prod_{i=1}^N\frac1{c_Na_i(r,\zeta)}
 \mathbb E_{r,\zeta}[F(p)],
\]
where $\mathbb E_{r,\zeta}$ is an expectation with respect all $P_i^{(r,\zeta)}$.
Since \(a_i(\zeta,r)\) is generally complex, this is formal notation rather than an actual probability expectation.

After absorbing  $c_N^{-N}$ into the normalization, this gives
\begin{align}\label{factored-positive-orthant}
&\int F(p)d\mu_{N,E}^{KR}(p)\\ \notag
&\propto\int_{\mathbb R^2}\int_{\mathbb R} \int_{\Gamma}  e^{c_N\Xi_N^{(1)}(r,\zeta,y)} (y_1-y_2)^2\mathbb E_{r,\zeta}\left[F(p)\prod\limits_{\alpha=1}^2h(\bar p,y_\alpha)\right] dy_1 dy_2 dr d\zeta
\end{align}
with
\begin{equation}\label{Xi_1}
\Xi_N^{(1)}(r,\zeta, y)=\cF_N(r,\zeta)+\cL_N(y_1)+\cL_N(y_2).
\end{equation}
Our aim is to prove that for the specific classes of observables we are interested in the leading order of the integral can be evaluated by
computing  $\mathbb E_{r,\zeta}$ and $h(\bar p,y_\alpha)$ at the saddle-points of $\cF_N(r,\zeta)$ and $\cL(y)$.

\section{Saddle point analysis}\label{s:2}
The saddle estimates below use only the local regularity event
$\Omega_N^{\mathrm{loc}}$. In the proofs of the limit theorems, we
work on the event $\Omega_N=\Omega_N^{\mathrm{loc}}\cap\Omega_N^{(2)}$.
\subsection{Saddle points}
The saddle equations for $\mathcal F_N$ from  \eqref{F} are
\begin{equation}\label{scalar-saddle-equations}
 1=c_N^{-1}\sum_j\frac{1}{(\tilde b_j-r/2)^2+\zeta},
 \qquad
 r/2=c_N^{-1}\sum_j\frac{\tilde b_j-r/2}{(\tilde b_j-r/2)^2+\zeta}.
\end{equation}
By \eqref{of:pastur}, it has a real solution 
\begin{equation}\label{r,zeta_*}
 r_*=-2t_N,\qquad \zeta_*=\eta_N^2.
\end{equation}
Take now $\zeta$ contour to be $\zeta_*+i\mathbb R$. Notice that since $\zeta_*>0$, we have $\Re a_i(r,\zeta)>0$, and so
the $p$-integral is absolutely convergent.

The saddle equation for $\mathcal L_N$ from (\ref{Phi,L,h}) is
\[
c_N^{-1}\sum_i\frac{i}{\tilde b_i+iy}=y,
\]
and, again by \eqref{of:pastur}, the complex solutions are $\pm \eta_N-it_N.$ 
One can easily check that $y$-contours can be moved to $-it_N+\mathbb R$, so the saddle points for $y$ become 
\begin{equation}\label{y_*}
y_\pm=\pm\eta_N.
\end{equation}
It is also useful to define
\begin{equation}
\widehat{\mathcal L}_N(y):=\mathcal L_N(y-it_N),\qquad
\widehat h(p,y):=h(p,y-it_N)
\end{equation}
and redefine $\Xi_N^{(1)}$ of (\ref{Xi_1}) as
\begin{equation}\label{Xi}
\Xi_N(r,\zeta,y)=\mathcal F_N(r,\zeta)+\widehat{\mathcal L}_N(y_1)
+\widehat{\mathcal L}_N(y_2).
\end{equation}

\begin{lemma}\label{l:cont}
On the contours $\zeta\in \{\zeta_*+it: t\in\mathbb R\}$, $r\in \mathbb R$, the function $\Re \cF(\zeta,r)$ has a unique global maximum at $r=r_*$, $t=0$.

Similarly, the function $\Re \widehat\cL_N(y)$, $y\in\mathbb R$ has global maxima attained precisely at $y=y_{\pm}$. 
\end{lemma}
\begin{proof}
Since $(\zeta_*,r_*)$ is the solution of (\ref{scalar-saddle-equations}), it is an extremum point of $\Re \cF(\zeta,r)$ (we will check that it is maximum below in Lemma \ref{l:F,L}).

Now let us consider
\[
\Re\cF(\zeta_*+it, r)-\Re\cF(\zeta_*, r_*)=-\dfrac{r^2}{4}+t_N^2-\dfrac{1}{2c_N}\sum_i\log \dfrac{((\tilde b_i-r/2)^2+\eta_N^2)^2+t^2}{(\beta_i^2+\eta_N^2)^2},
\]
where $\beta_i=\tilde b_i+t_N(E)$.

For a fixed $r$, this function has a maximum at $t=0$, so we can study
\begin{equation}
F_N(r)=-\dfrac{r^2}{4}+t_N^2-\dfrac{1}{c_N}\sum_i\log \dfrac{(\tilde b_i-r/2)^2+\eta_N^2}{\beta_i^2+\eta_N^2}
\end{equation}
It is easy to see
\[
F''_N(r)=-\dfrac{1}{2}+\dfrac{1}{2c_N}\sum_i \dfrac{1}{(\tilde b_i-r/2)^2+\eta_N^2}-\dfrac{1}{c_N}\sum_i \dfrac{\eta_N^2}{((\tilde b_i-r/2)^2+\eta_N^2)^2}.
\]
Notice that for any fixed $R>0$ Proposition \ref{p:LLN} and (\ref{of:L1}), (\ref{of:L2-scaled}) imply 
\[
\sup\limits_{|r|\le R,c\tau_N^2<\eta\le C\tau_N^2}\Big|\dfrac{1}{2c_N}\sum_i \dfrac{1}{(\tilde b_i-r/2)^2+\eta^2}-\dfrac{1}{c_N}\sum_i \dfrac{\eta^2}{\big((\tilde b_i-r/2)^2+\eta^2\big)^2}\Big|\le O(\tau_N^2+\varepsilon_N),
\]
and so $$F''_N(r)=-1/2+o(1)<0$$ for $|r|<R$, and hence $F_N(r)$ has a unique maximum on $|r|<R$ for any fixed $R$.

It remains to control $|r|>R$ for big enough $R$. Using the
elementary inequality
\[
 \frac{x^2+\eta^2}{(x-h)^2+\eta^2}
 \leq2\left(1+\frac{h^2}{\eta^2}\right),
 \qquad x,h\in\mathbb R,
\]
applied with $x=\beta_i$ and $h=r/2+t_N$, we get
\[
F_N(r)\le -\dfrac{r^2}{4}+t_N^2+\tau^2_N\log \Big[2\Big(1+\frac{(r/2+t_N)^2}{\eta_N^2}\Big)\Big].
\]
Because $\eta_N\asymp\tau_N^2$, one may choose a deterministic $R$ and
then $N$ sufficiently large so that the right-hand side is at most
$-c r^2$ for every $|r|\geq R$. This finishes the proof of the first statement of the lemma.

To check the statement about $\Re \widehat\cL_N(y)$, $y\in\mathbb R$, consider
\[
\Re \widehat\cL_N'(y)=y\left(c_N^{-1}\sum_i\dfrac{1}{\beta_i^2+y^2}-1\right).
\] 
Since the function in parentheses is monotonic in $y$, the extremums for $y>0$ and for $y<0$ are unique. 
\end{proof}

Denote $F|_*$ the value of a function at a saddle point $(r_*,\zeta_*, y_*)$ where $y_*$ means one of the points $y_\pm$.
We will need below
\begin{lemma}\label{l:F,L}
In the notations above we have
\begin{align*}
&-\dfrac{\partial^2 \cF_N}{\partial \zeta^2}\Big|_*= -\frac{\tau_N^{-4}}{2(\pi\rho(E))^2}(1+o(1)),\quad \dfrac{\partial^2 \cF_N}{\partial r^2}\Big|_*=-1/2+o(1),\\
& \dfrac{\partial^2 \cF_N}{\partial \zeta \partial r}\Big|_*=o(\tau_N^{-2}),\quad \dfrac{\partial^2  \widehat\cL_N(y)}{\partial y^2}\Big|_*=-1+o(1).
\end{align*}
In addition, for $\sigma_1,\sigma_2\in\{+1,-1\}$, make the change
\begin{equation}\label{change_z,r,y}
\zeta=\zeta_*+\dfrac{i\tau_N^3}{\sqrt{N}}\tilde\zeta,\quad r=r_*+\dfrac{\tau_N}{\sqrt{N}}\tilde r, \quad y_\alpha=\sigma_\alpha\eta_N+\dfrac{\tau_N}{\sqrt{N}}\tilde y_\alpha,\quad \alpha=1,2,
\end{equation}
and set
\begin{align}
U_N^{\sigma_1,\sigma_2}:={}&\{(\zeta,r,y_1,y_2):|\tilde\zeta|+|\tilde r|+|\tilde y_1|+|\tilde y_2|\le \log N\},
\qquad
U_N:=\bigcup_{\sigma_1,\sigma_2\in\{+1,-1\}}U_N^{\sigma_1,\sigma_2}.
\label{Om}
\end{align}
then uniformly in $U_N$ we have
\begin{align}\label{third_der}
&\left|\dfrac{\partial^3 \cF_N}{\partial \tilde\zeta^3}\right|+
\left|\dfrac{\partial^3 \cF_N}{\partial \tilde\zeta^2\partial \tilde r}\right|+
\left|\dfrac{\partial^3 \cF_N}{\partial \tilde\zeta\partial\tilde r^2}\right|+
\left|\dfrac{\partial^3 \cF_N}{\partial \tilde r^3}\right|+
\left|\dfrac{\partial^3 \widehat\cL_N}{\partial \tilde y^3}\right|
\le C\frac{\tau_N}{N^{3/2}}.
\end{align}
Moreover, in $(\tilde\zeta,\tilde r, \tilde y_1, \tilde y_2)$ variables we have
\begin{align}\label{Hess}
&F_*=-c_N\text{Hess}(\cF_N)_{\tilde\zeta,\tilde r}\Big|_*=\left(\begin{matrix}
c_0+o(1)&o(1)\\
o(1)& 1/2+o(1)
\end{matrix}\right)\Longrightarrow F_*^{-1}=\left(\begin{matrix}
c_0^{-1}+o(1)&o(1)\\
o(1)& 2+o(1)
\end{matrix}\right)\notag\\
&L_*=-c_N(\widehat\cL_N^{\prime\prime})_{\tilde y}\Big|_*=1+o(1)\Longrightarrow L_*^{-1}=1+o(1)
\notag\end{align}
with $c_0=\frac{1}{2(\pi\rho(E))^2}$. Note also that in $U_N$ 
\begin{equation}\label{vander}
(y_1-y_2)^2=\begin{cases}
4\eta_N^2(1+O((\tau_N^2N)^{-1/2}\log N)),& y_1=\pm \eta_N+\dfrac{\tau_N}{\sqrt{N}}\tilde y_1,\quad  y_2=\mp \eta_N+\dfrac{\tau_N}{\sqrt{N}}\tilde y_2,\\
\eta_N^2 \cdot O((\tau_N^2N)^{-1}\log^2 N)& y_1=\pm \eta_N+\dfrac{\tau_N}{\sqrt{N}}\tilde y_1,\quad  y_2=\pm \eta_N+\dfrac{\tau_N}{\sqrt{N}}\tilde y_2.
\end{cases}
\end{equation}
and therefore for $\Xi_N$ of (\ref{Xi}) we get an expansion
\begin{equation}\label{exp_Xi}
c_N(\Xi_N-\Xi_{N,*})=-\frac12(F_*\Delta_1,\Delta_1)-\frac12L_*(\Delta_2,\Delta_2)+O(\log^3N(N\tau_N^2)^{-1/2}),
\end{equation}
where $\Delta_1=(\tilde \zeta,\tilde r)^t$, $\Delta_2=(\tilde y_1,\tilde y_2)^t$.

%
\end{lemma}
\begin{proof}
Denote $\beta_i=\tilde b_i+t_N$. Computing the derivatives and using Proposition \ref{p:LLN} and Lemma \ref{l:int} we get
\begin{align*}
&\dfrac{\partial^2 \cF_N}{\partial \zeta^2}\Big|_*=\dfrac{\tau_N^2}{N}\sum\limits_i\dfrac{1}{a_i(r,\zeta)^2}\Big|_*=\dfrac{\tau_N^2}{N}\sum\limits_i\dfrac{1}{(\beta_i^2+\eta^2_N)^2}=\frac{\tfrac12+o(1)}{\eta_N^2}\propto \tau_N^{-4},\\
&\dfrac{\partial^2 \cF_N}{\partial r^2}\Big|_*=\dfrac{\tau_N^2}{N}\sum\limits_i\dfrac{\beta_i^2}{(\beta_i^2+\eta^2_N)^2}-1=-1/2+o(1),\\
&\dfrac{\partial^2 \cF_N}{\partial\zeta\partial r}\Big|_*=-\dfrac{\tau_N^2}{N}\sum\limits_i\dfrac{\beta_i}{(\beta_i^2+\eta^2_N)^2}=o(\tau_N^{-2}),\\
&\dfrac{\partial^2  \widehat\cL_N(y)}{\partial y^2}\Big|_*=-\dfrac{\tau_N^2}{N}\sum\limits_i\dfrac{2\eta_N^2}{(\beta_i^2+\eta^2_N)^2}\pm\dfrac{i\tau_N^2}{N}\sum\limits_i\dfrac{2\eta_N\beta_i}{(\beta_i^2+\eta^2_N)^2}=-1+o(1),
\end{align*}
which proves the first statement and the expression for the Hessian after the change of variables. The third derivatives are
\begin{align*}
&\dfrac{\partial^3 \cF_N}{\partial \zeta^3}=-\dfrac{\tau_N^2}{N}\sum\limits_i\dfrac{2}{a_i(r,\zeta)^3}\\
&\dfrac{\partial^3 \cF_N}{\partial r^3}=-\dfrac{3\tau_N^2}{2N}\sum\limits_i\dfrac{(\tilde b_i-r/2)}{a_i(r,\zeta)^2}+\dfrac{2\tau_N^2}{N}\sum\limits_i\dfrac{(\tilde b_i-r/2)^3}{a_i(r,\zeta)^3},\\
&\dfrac{\partial^3 \cF_N}{\partial\zeta^2\partial r}=\dfrac{\tau_N^2}{N}\sum\limits_i\dfrac{2(\tilde b_i-r/2)}{a_i(r,\zeta)^3},\\
&\dfrac{\partial^3 \cF_N}{\partial\zeta\partial r^2}=\dfrac{\tau_N^2}{2N}\sum\limits_i\dfrac{1}{a_i(r,\zeta)^2}-\dfrac{2\tau_N^2}{N}\sum\limits_i\dfrac{(\tilde b_i-r/2)^2}{a_i(r,\zeta)^3},\\
&\dfrac{\partial^3  \Re \widehat\cL_N(y)}{\partial y^3}=-\dfrac{\tau_N^2}{N}\sum\limits_i\dfrac{6y}{((\tilde b_i+t_N)^2+y^2)^2}+
\dfrac{\tau_N^2}{N}\sum\limits_i\dfrac{8y^3}{((\tilde b_i+t_N)^2+y^2)^3},\\
&\dfrac{\partial^3  \Im \widehat\cL_N(y)}{\partial y^3}=-\dfrac{2\tau_N^2}{N}\sum\limits_i\dfrac{(\tilde b_i+t_N)}{((\tilde b_i+t_N)^2+y^2)^2}+
\dfrac{\tau_N^2}{N}\sum\limits_i\dfrac{8(\tilde b_i+t_N)y^2}{((\tilde b_i+t_N)^2+y^2)^3}
\end{align*}
Using
\[
\Big|\dfrac{1}{a_i(r,\zeta)}\Big|\le \dfrac{1}{\eta_N^2},\quad \Big|\dfrac{(\tilde b_i-r/2)}{a_i(r,\zeta)}\Big|\le \dfrac{|\tilde b_i-r/2|}{(\tilde b_i-r/2)^2+\eta_N^2}\le \dfrac{2}{\eta_N},
\]
similar inequalities with $y$, and Proposition \ref{p:LLN}, we obtain 
\begin{align*}
&\dfrac{\tau_N^2}{N}\sum\limits_i\Big|\dfrac{1}{a_i(r,\zeta)^3}\Big|\le \eta_N^{-4}\cdot \dfrac{\tau_N^2}{N}\sum\limits_i\dfrac{1}{(\tilde b_i-r/2)^2+\eta_N^2}=O(\eta_N^{-4})=O(\tau_N^{-8}),\\
&\dfrac{\tau_N^2}{N}\sum\limits_i\Big|\dfrac{(\tilde b_i-r/2)}{a_i(r,\zeta)^2}\Big|\le \dfrac{2}{\eta_N}\cdot \dfrac{\tau_N^2}{N}\sum\limits_i\dfrac{1}{(\tilde b_i-r/2)^2+\eta_N^2}=O(\eta_N^{-1})=O(\tau_N^{-2}).
\end{align*}
Other terms can be bounded similarly. Changing to $(\tilde\zeta,\tilde r, \tilde y_1, \tilde y_2)$ variables, we get (\ref{third_der}).
\end{proof}

\subsection{Proofs of Theorems \ref{t:IPR}-\ref{t:min}}

\textit{Proof of Theorem \ref{t:IPR}}. It is easy to see that to prove (\ref{IPR}), it is sufficient to prove that
\begin{align}\label{E,Var}
\int I_q^{s}\,d\mu^{KR}_{N,E} =\Big(\Gamma(q+1)\sum_{i=1}^Nm_i(E)^q\Big)^{s}(1+o(1)),\quad s=1,2
\end{align}
where $\mu^{KR}_{N,E} $ was defined in (\ref{mu^KR}).
Denote
\begin{align}\label{<>}
&\langle\dots\rangle=\int (\dots)(y_1-y_2)^2\exp\{c_N \Xi_N(\zeta,r,y)\}d\zeta dr dy_1dy_2 
\end{align}
where $\Xi(\zeta,r,y)$ is defined in \eqref{Xi}.


Let $\mathcal{Z}$ be a normalizing factor in (\ref{mu^KR}), i.e.
\begin{align}\label{Z}
\mathcal{Z}=&\int dpd\zeta dr dy_1dy_2 \prod\limits_{l=1}^2\widehat h(\bar p,y_l)\\
&\times e^{c_N(\widehat\cL_N(y_1)+\widehat \cL_N(y_2)+\Phi_N(\bar p,\zeta,r))} (y_1-y_2)^2.
\end{align}
Integrating  with respect to $d\bar p$, we get
\begin{align*}
\mathcal{Z}
=&\langle g(\zeta,r,y_1)g(\zeta,r,y_2)+\Sigma'\rangle \notag
\end{align*}
where
\begin{align}
&g(\zeta,r,y)=c_N^{-1}\sum_{j}
\frac{(\tilde b_{j}+t_N+iy)^{-1}}{(\tilde b_{j}-r/2)^2+\zeta}\label{g}\\
&\Sigma'= c_N^{-2}\sum_j\frac{(\tilde b_{j}+t_N+iy_1)^{-1}(\tilde b_{j}+t_N+iy_2)^{-1}}{((\tilde b_{j}-r/2)^2+\zeta)^2}
\notag\end{align}
Notice that due to Lemmas \ref{l:cont} -- \ref{l:F,L} by the standard saddle-point analysis argument 
the contribution of the $U_N^{c}$ to the integral is bounded by $e^{-C\log^{2}N}$. Indeed, 
in each of the four saddle neighbourhoods in (\ref{Om}), Taylor expansion and (\ref{third_der}) give
\[
c_N\Re\bigl(\Xi_N-\Xi_{N,*}\bigr)
\leq-c\bigl(\tilde\zeta^2+\tilde r^2+\tilde y_1^2+\tilde y_2^2\bigr)
+O\!\left((N\tau_N^2)^{-1/2}
\bigl(|\tilde\zeta|+|\tilde r|+|\tilde y_1|+|\tilde y_2|\bigr)^3\right).
\]
Hence, on the boundary of $U_N$, the real part of the exponent is smaller than its saddle value by $c\log^2N$.  Lemma~\ref{l:cont} propagates this gap outside a fixed neighbourhood: the real part decreases in $|\Im\zeta|$, strict concavity controls the $r$-direction, and the monotonicity assertion for $\Re\widehat\cL_N(y)$ controls the two $y$-directions.  The explicit tail bounds in the proof of that lemma treat the unbounded portions of the contours.  Consequently, the contribution of $U_N^c$ to the saddle integral with any polynomially bounded prefactor is $O(e^{-c\log^2N})$.


In addition, by using Lemmas \ref{l:int}-\ref{l:sol} and Proposition \ref{p:LLN} it is easy to get w.h.p that inside $U_N$ of (\ref{Om})
\begin{align}\label{b_g}
|g(\zeta_*,r_*,y_*)|\asymp \eta_N^{-1},\quad |\Sigma'|\le C/N\eta_N^3,
\end{align}
so
\begin{align*}
|\Sigma'|\le C|g(\zeta_*,r_*,y_*)|^2(\tau_N^2N)^{-1}\Rightarrow 
\mathcal{Z}=\langle g(\zeta,r,y_1)g(\zeta,r,y_2)\rangle(1+o(1))
\end{align*}
Similarly, integrating by $\bar p$ in (\ref{E,Var}) we get
\begin{align*}
\int I_q^{s}\,d\mu^{KR}_{N,E}=\frac{\langle f_q^s(\zeta,r)g(\zeta,r,y_1)g(\zeta,r,y_2)\rangle}{\langle g(\zeta,r,y_1)g(\zeta,r,y_2)\rangle}(1+O((\tau_N^2N)^{-1})),\,s=1,2,
\end{align*}
where
\begin{align}\label{f_q}
f_q(\zeta,r)=\Gamma (q+1)c_N^{-q}\sum\limits_j \frac{1}{((\tilde b_{j}-r/2)^2+\zeta)^q}.
\end{align}

Let  us prove the following simple lemma:
\begin{lemma}\label{l:1} Let the function $f(\zeta,r,y)$ uniformly  in $U_N$ satisfies the relations:
\begin{align}\label{cond_0}
f= f_*(1+o(1)), \quad f_*=f(\zeta,r,y)\Big|_*
\end{align}
Then
\begin{align}\label{l1.2}
\langle f\rangle=\langle f_*\rangle(1+o(1)).
\end{align}
The sufficient conditions for (\ref{cond_0}) are
\begin{align}\label{cond_f}
&  \Big|\frac{\partial f}{\partial\zeta }\Big|\le |f_*|\tau_N^{-4}\nu_N,\quad  \Big|\frac{\partial f}{\partial r }\Big|\le |f_*|\tau_N^{-2}\nu_N,
\quad  \Big|\frac{\partial f}{\partial y_{1,2} }\Big|\le |f_*|\tau_N^{-2}\nu_N
\end{align}
uniformly in $U_N$, where $\{\nu_N\}$ is some sequence growing slowly than any degree of $N$. 

\end{lemma}

\textit{Proof.} The statement of the lemma is trivial for positive measures. However, since the measure $<...>$ is not positive, we need to show the relation 
\begin{align*}
&\frac{\int\exp\{c_N\Re \Xi_N\}|y_1-y_2|^2d\zeta dr dy|}
{\Big|\int\exp\{c_N\Xi_N\}(y_1-y_2)^2d\zeta dr dy\Big|}=
\frac{ |(\Re L_*)^2\det \Re F_* |^{-1/2}}{|L_*^2\det  F_* |^{-1/2}}(1+o(1)) \le C,
\end{align*}
if we choose the same $y_\pm$ for $y_1$ and $y_2$ in $U_N$, and 
\begin{align*}
&\frac{\int\exp\{c_N\Re \Xi_N\}d\zeta dr dy|}
{\Big|\int\exp\{c_N\Xi_N\}d\zeta dr dy\Big|}=
\frac{|(\Re L_*)^4\det \Re F_* |^{-1/2}}{  |L_*^4\det  F_* |^{-1/2}}(1+o(1)) \le C,
\end{align*}
if we choose different $y_\pm$  for $y_1$ and $y_2$ (see (\ref{vander})).

Note that both inequalities easily follow from (\ref{Xi}), (\ref{exp_Xi}), and  saddle point computations of Lemma \ref{l:F,L}.

To check that (\ref{cond_f}) implies (\ref{cond_0}) we  use  the change (\ref{Om}) for  $f$. Indeed, denoting $\nabla f$ the vector of  derivatives
of $f$  with respect to the variable $(\tilde\zeta,\tilde r,\tilde y_1,\tilde y_2)$ we get from (\ref{cond_f}) 
\[
|\nabla f|\le C| f_* |\nu_N/(\tau_N^2N)^{1/2},
\]
which implies (\ref{cond_0}) after the trivial Taylor expansion.

%

$\square$

Using the lemma, we conclude that to finish the proof of Theorem \ref{t:IPR}, it suffices to check conditions (\ref{cond_f})
for $f=g(\zeta,r,y_1)g(\zeta,r,y_2)$ and for $f=f_q^s(\zeta,r)g(\zeta,r,y_1)g(\zeta,r,y_2)$, $s=1,2$. 

We first verify the saddle-neighbourhood estimates for $f_q$ directly.
Set
\[
 a_j(r,\zeta)=(\widetilde b_j-r/2)^2+\zeta,
 \qquad
 a_{j,*}=\beta_j^2+\eta_N^2,
 \qquad
 \beta_j=\widetilde b_j+t_N.
\]
Writing $\Delta r=r-r_*$ and $\Delta\zeta=\zeta-\eta_N^2$, we have
\[
 a_j(r,\zeta)-a_{j,*}
 =-\beta_j\Delta r+\frac{(\Delta r)^2}{4}+\Delta\zeta.
\]
Since
\[
 \frac{|\beta_j|}{\beta_j^2+\eta_N^2}\le \frac1{2\eta_N},
 \qquad
 \frac1{\beta_j^2+\eta_N^2}\le \eta_N^{-2},
\]
the definition of $U_N$ implies, uniformly in $j$ and in $U_N$,
\[
 \left|
 \frac{a_j(r,\zeta)-a_{j,*}}{a_{j,*}}
 \right|
 \le C\left(
 \frac{|\Delta r|}{\eta_N}
 +\frac{|\Delta r|^2+|\Delta\zeta|}{\eta_N^2}
 \right)
 \le C\frac{\log N}{\sqrt{N\tau_N^2}}
 =:d_N=o(1).
\]
Thus, for every fixed $q>0$,
\[
 a_j(r,\zeta)^{-q}
 =a_{j,*}^{-q}\bigl(1+O_q(d_N)\bigr)
\]
uniformly in $j$ and in $U_N$. Hence,
\[
 f_q(r,\zeta)=f_{q,*}\bigl(1+O_q(d_N)\bigr),
 \qquad
 f_{q,*}=\Gamma(q+1)c_N^{-q}\sum_j a_{j,*}^{-q}>0.
\]
The same comparison yields
\[
 |\partial_\zeta f_q|\le C_q\eta_N^{-2}f_{q,*},
 \qquad
 |\partial_r f_q|\le C_q\eta_N^{-1}f_{q,*}.
\]
Together with the corresponding estimates for $g$, these bounds verify
the conditions of Lemma~\ref{l:1} for
\[
 g(\zeta,r,y_1)g(\zeta,r,y_2),
 \qquad
 f_q(r,\zeta)^s
 g(\zeta,r,y_1)g(\zeta,r,y_2),
 \quad s=1,2.
\]

Note also that because of (\ref{vander}), the leading contribution is given by the saddle-points $y_{1,*}=\pm \eta_N,\quad  y_{2,*}=\mp \eta_N$.
Indeed the $\langle..\rangle$ measure contains $(y_1-y_2)^2$.  It is of
order $\eta_N^2$ for $y_{1,*}=\pm \eta_N,\quad  y_{2,*}=\mp \eta_N$ and of order $\tau_N^2/N$ for
$y_{1,*}=\pm \eta_N,\quad  y_{2,*}=\pm \eta_N$ on the Gaussian fluctuation scale, and the other functions with respect to $y$ (i.e., $h$, $\cL$) just have a complex conjugated values at $y=\pm \eta_N$.  Therefore, the total
same-sign contribution is $O(K_N^{-1})$ relative to the mixed-sign one.

It remains to check (\ref{IPR-supercritical}) -- (\ref{IPR-subcritical}). 
If $q>1/2$,  then  (\ref{LLN4}) for $F_q(u)=(1+u^2)^{-q}$ yields
\[
 K_N^{q-1}\sum_i m_i^q
 =\frac1{K_N}\sum_iF_q(\tfrac{\tilde b_i+t_N}{\eta_N})=\frac{N}{K_N}\Big(\int F_q(\tfrac{b-E+t_N}{\eta_N}) \rho(b) db+o(\eta_N)\Big).
 \]
Changing $u=\tfrac{b-E+t_N}{\eta_N}$, we get
\[
 \eta_N(E)\int F(u)\rho(E-t_N(E)+\eta_N(E)u)\,du
 =\eta_N(E)\rho(E)\int F(u)\,du\,[1+o(1)]
\]
uniformly in $E$.  Since
\[
 \frac{N\eta_N\rho(E)}{K_N(E)}
 =\frac{\tau_N^2\rho(E)}{\eta_N}=\frac1\pi+o(1),
\]
we get
\[
 K_N^{q-1}\sum_i m_i^q
  \longrightarrow \frac1\pi\int_{\mathbb R}\frac{du}{(1+u^2)^q}
 =\frac{\Gamma(q-1/2)}{\sqrt\pi\,\Gamma(q)}.
\]
This proves \eqref{IPR-supercritical}.

For $0<q<1/2$, fix $h>0$ and split at $|b_i-E|=h$.  On the complement the
function $|x-E|^{-2q}$ is bounded and we get similarly
\[
 N^{\gamma q-1}\sum_{|b_i-E|>h}m_i^q
 \longrightarrow\int_{|x-E|>h}|x-E|^{-2q}\rho(x)\,dx.
\]
The remaining part is bounded, uniformly in $N$, by
$C\int_{|x-E|\leq 2h}|x-E|^{-2q}\rho(x)dx+o(1)$, which tends to zero as
$h\downarrow0$.  This proves \eqref{IPR-subcritical}.

For $q=1/2$, the $s=1/2$ case of \eqref{LLN2}, applied at
$r=-2t_N$ and $\zeta=\eta_N^2$, gives
\[
 \frac1N\sum_{j=1}^N
 \frac1{\sqrt{(b_j-E+t_N)^2+\eta_N^2}}
 =
 \int_{\mathbb R}
 \frac{\rho(b)\,db}
 {\sqrt{(b-E+t_N)^2+\eta_N^2}}
 +O(\varepsilon_N).
\]
By the smoothness of $\rho$ near $I$ and the fact that $t_N=o(1)$,
uniformly in $E\in I$,
\[
 \int_{\mathbb R}
 \frac{\rho(b)\,db}
 {\sqrt{(b-E+t_N)^2+\eta_N^2}}
 =
 2\rho(E)\log\frac1{\eta_N}
 +o\!\left(\log\frac1{\eta_N}\right).
\]
Since
\[
 K_N^{-1/2}\sum_{j=1}^N m_j^{1/2}
 =
 \frac{\tau_N^2}{N\eta_N}
 \sum_{j=1}^N
 \frac1{\sqrt{(b_j-E+t_N)^2+\eta_N^2}},
 \qquad
 \frac{\tau_N^2\rho(E)}{\eta_N}
 =\frac1\pi[1+o(1)],
\]
we obtain
\[
 K_N^{-1/2}\sum_{j=1}^N m_j^{1/2}
 =
 \frac2\pi\log\frac1{\eta_N}\,[1+o(1)].
\]
This proves \eqref{IPR-critical}.


$\square$

\medskip

\textit{Proof of Theorem \ref{thm:ampl}.}  It is easy to see that  to  prove the theorem it is sufficient 
to prove relations
\begin{align}\label{N(a).1}
&\mathbb{E}\big[N^{-f(\alpha)}\mathcal{N}_{N}(\alpha)\big]= \sqrt \pi\rho (E)+o(1),\quad 2-\gamma<\alpha<\gamma,\\
&\mathbb{E}\big[N^{-f(\alpha)}\mathcal{N}_{N}(\alpha)\big]=\int \rho(b)e^{-(b-E)^2}db+o(1),\quad \alpha=\gamma, \notag \\
&\mathbb{E}\big[N^{-f(\alpha)}(N-\mathcal{N}_{N}(\alpha))\big]= \mu_2(E)+o(1),\quad  \gamma+1>\alpha>\gamma, \notag\\
&\mathrm{Var}\big[\mathcal{N}_{N}(\alpha)N^{-f(\alpha)}\big]=o(1),\notag
\end{align}
for  
\[ \mathcal N_N(\alpha)=\#\{i:
  p_i\geq N^{-\alpha}\}.\]
Here $\mu_2(E)$ is  defined in (\ref{m_2}).  
  
Indeed,  let $2-\gamma<\alpha<\gamma$ and $\varepsilon$ is small enough. 
Write
\begin{align}\label{log}
\frac{\log\mathcal N_N(\alpha,\varepsilon)}{\log N}=\frac{\log\Big(\mathcal N_N(\alpha+\varepsilon)-\mathcal N_N(\alpha-\varepsilon)\Big)}{\log N}=
\frac{\log\mathcal N_N(\alpha+\varepsilon)+\log\Big(1-\tfrac{\mathcal N_N(\alpha-\varepsilon)}{\mathcal N_N(\alpha+\varepsilon)}\Big)}{\log N}.
\end{align}
Fix  small $\delta>0$.  The relation (\ref{N(a).1}) implies
\begin{align*}
\mathcal{N}_{N}(\alpha\pm\varepsilon)/N^{f(\alpha\pm\varepsilon)}\xrightarrow{p} &\sqrt \pi\rho (E)\\
\Longrightarrow & \p\Big(\tfrac{\log(\sqrt{\pi}\rho (E)-\delta)}{\log N}\le 
\tfrac{\log \mathcal{N}_{N}(\alpha+\varepsilon)}{\log N}-f(\alpha+\varepsilon)\le \tfrac{\log(\sqrt{\pi}\rho (E)+\delta)}{\log N}\Big)\to 1,
\end{align*} 
and so the first term of (\ref{log}) goes to $f(\alpha+\varepsilon)$, as $N\to \infty$.
Similarly, since $f(\alpha+\varepsilon)-f(\alpha-\varepsilon)=\varepsilon$,
\[
\p\Big(\tfrac{1-\delta}{N^{f(\alpha+\varepsilon)-f(\alpha-\varepsilon)}}\le \tfrac{\mathcal N_N(\alpha-\varepsilon)}{\mathcal N_N(\alpha+\varepsilon)}\le \tfrac{1+\delta}{N^{f(\alpha+\varepsilon)-f(\alpha-\varepsilon)}}\Big)\to 1
\]
 implies that the second term of (\ref{log}) goes to $0$, as $N\to\infty$. The case $\alpha> \gamma$ can be considered in a similar way using
 $C_N(\alpha)=N-\mathcal{N}_{N}(\alpha)$ instead of $\mathcal N_N(\alpha)$.
 For $\alpha=\gamma$, the estimates on the two branches give, for
sufficiently small $\varepsilon>0$,
\[
 \mathcal N_N(\gamma-\varepsilon)=o_{\mathbb P}(N),
 \qquad
 C_N(\gamma+\varepsilon)=o_{\mathbb P}(N).
\]
Hence,
\[
 \mathcal N_N(\gamma,\varepsilon)
 =N-C_N(\gamma+\varepsilon)-\mathcal N_N(\gamma-\varepsilon)
 =N[1+o_{\mathbb P}(1)],
\]
which yields $f(\gamma)=1$.
  
To prove  the first  relation of (\ref{N(a).1}),  observe that for different $j_1,j_2,j_3$ 
\begin{align}\label{sigma}
&N^{-f(\alpha)}\int  d\bar p e^{N\tau_N^{-2} \Phi_N(\bar p,\zeta,r)}\frac{p_{j_1}}{\tilde b_{j_1}+iy_1}\frac{p_{j_2}}{\tilde b_{j_2}+iy_2}\theta(p_{j_3}-N^{-\alpha}) \\=
&c_N^{-2}N^{-f(\alpha)}\frac{((\tilde b_{j_1}-r/2)^2+\zeta)^{-1}}{\tilde b_{j_1}+iy_1}\frac{((\tilde b_{j_2}-r/2)^2+\zeta)^{-1}}{\tilde b_{j_2}+iy_2}e^{-N^{\gamma-\alpha}((\tilde b_{j_3}-r/2)^2+\zeta)}
\int  d\bar p e^{N\tau_N^{-2} \Phi_N(\bar p,\zeta,r)}
\notag\end{align}
Introduce the function
\begin{align}
V_\alpha(\zeta,r)=N^{-f(\alpha)}\sum_j e^{-N^{\gamma-\alpha}((\tilde b_{j}-r/2)^2+\zeta)}.
\label{V(a)}\end{align}
For $2-\gamma<\alpha< \gamma$, by (\ref{LLN3})  in $\Omega_N$ we have uniformly in $U_N$
\begin{align*}
&V_\alpha(\zeta,r)=N^{1-f(\alpha)}\int \rho (b) db e^{-N^{\gamma-\alpha}((b-E-r/2)^2+\zeta)}+O(N^{-1/2+(\gamma-\alpha)/4}\log N)\\
&=\sqrt\pi\rho(E+r/2)N^{1-(\gamma-\alpha)/2-f(\alpha)}
\big(1+O(N^{2-\gamma-\alpha})+O(N^{-(\gamma-\alpha)})+O(N^{-1/2+(\gamma-\alpha)/4}\log N)\big)\\
&=\sqrt\pi\rho(E)(1+O(r)+O(N^{2-\gamma-\alpha})+O(N^{-(\gamma-\alpha)}))+O(N^{-1/2+(\gamma-\alpha)/4}\log N)\\
&=V_{\alpha,*}(1+o(1)).
\end{align*}
Then the sum of terms of (\ref{sigma}) with $j_3=j_1$ or $j_3=j_2$ can be estimated as 
$N^{-f(\alpha)}g_*^2\asymp N^{-f(\alpha)}g_*^2 V_{\alpha,*}$ and the sum of terms with $j_3\not=j_1$ and $j_3\not=j_2$ was
estimated  in (\ref{b_g}). Thus the sum of integrals with some coinciding indices is bounded by $C(N^{-f(\alpha)}+(N\tau_N^2)^{-1}g_*^2 V_{\alpha,*})$.

Hence we get   for $2-\gamma<\alpha< \gamma$:
\begin{align*}\notag
\int \mu^{KR}_{N,E}(\bar p) d\bar p  \big(N^{-f(\alpha)}\mathcal N_N(\alpha)\big)=&
\frac{\langle V_\alpha(\zeta,r)g(\zeta,r,y_1)g(\zeta,r,y_2)\rangle}{ \langle g(\zeta,r,y_1)g(\zeta,r,y_2)\rangle}(1+O((N\tau_N^2)^{-1})+O(N^{-f(\alpha)}))\\
&=V_{\alpha,*}(1+o(1))
\end{align*}
which  implies
the first relation of (\ref{N(a).1}) for $2-\gamma<\alpha< \gamma$.

For $\alpha=\gamma$, ($f(\alpha)=1$), according to (\ref{LLN5}) we have
\begin{align*}
&V_\alpha(\zeta,r)=\int \rho (b) db e^{-(b-E)^2}+O(N^{-1/2}\log N)+O(|r|)+O(|\zeta|).
\end{align*}
All other estimates are the same.

\smallskip

 For $\alpha> \gamma$ by the same argument we obtain an analogue of (\ref{V(a)})
for $N^{-f(\alpha)}(N-\mathcal N_N(\alpha))$ with
\begin{align*}
\widetilde V_\alpha(\zeta,r) &=N^{-f(\alpha)}\sum_j \Big(1-e^{-N^{\gamma-\alpha}((\tilde b_{j}-r/2)^2+\zeta)}\Big)\\
&=\frac1N\sum_j a_j(r,\zeta)\,
 \psi\!\left(N^{\gamma-\alpha}a_j(r,\zeta)\right),\\
&=\frac1N\sum_j a_j(r,\zeta)+o(1)
 =\mu_2(E)+o(1),\quad \psi(z)=\frac{1-e^{-z}}{z}
\end{align*}
uniformly on \(U_N\). Here the second equality follows by a standard truncation argument, using $\psi(z)\to1$ as $z\to0$ on bounded sets and (\ref{mu_2.1}) to control the tails.

To estimate the sum of integrals with $\theta(N^{-\alpha}-p_{j_3})$ instead of $\theta(p_{j_3}-N^{-\alpha})$ in (\ref{sigma}) with $j_3=j_1$ or $j_3=j_2=j_1$ we use the relation
\[
\begin{aligned}
&\big|\int_0^{N^{-\alpha}}p_j^s
 e^{-N^{\gamma}((\tilde b_{j}-r/2)^2+\zeta)p_j}\,dp_j\big|\\
&\hspace{2cm}\le N^{-s\alpha}
 \int_0^{N^{-\alpha}}
 e^{-N^{\gamma}p_j\Re ((\tilde b_{j}-r/2)^2+\zeta)}\,dp_j,
 \qquad s=1,2.
\end{aligned}
\]
Denote by $\Sigma_1'$  the sum over  $j_3=j_1$ and $j_2\not=j_1$, and by $\Sigma_2'$  the sum with $j_3=j_2=j_1$.
Using the previous inequalities and $g_*\asymp \eta_N^{-1}$, we obtain
\[
|\Sigma_1'|+|\Sigma_2'|\le CN^{1-\alpha-\gamma}|\tilde V_* g_*^2|=o(|\tilde V_* g_*^2|).
\]
The sum of the terms with $j_1=j_2$ and $j_3\not=j_1$ was  estimated in (\ref{b_g}). Thus we get
\begin{align*}
&\int \mu^{KR}_{N,E}(\bar p)\,d\bar p\,
 N^{-f(\alpha)}\bigl(N-\mathcal N_N(\alpha)\bigr)\\
&\quad=
\frac{\langle \tilde V_\alpha(\zeta,r)g(\zeta,r,y_1)g(\zeta,r,y_2)\rangle}
{\langle g(\zeta,r,y_1)g(\zeta,r,y_2)\rangle}
\bigl(1+O((N\tau_N^2)^{-1})+O(N^{1-\alpha-\gamma})\bigr)\\
&\quad=\tilde V_{\alpha,*}(1+o(1)).
\end{align*}
Thus we obtain the third relation in (\ref{N(a).1}).

To obtain the last relation of (\ref{N(a).1}) we apply the same argument to $V_\alpha^2$ and $\tilde V_\alpha^2$. The terms with two distinct marked indices give
\[
V_{\alpha,*}^{\,2}(1+o(1)),
\qquad\text{respectively}\qquad
\widetilde V_{\alpha,*}^{\,2}(1+o(1)),
\]
while the terms with coinciding indices are $o(1)$ by the estimates above. Thus the corresponding second moment equals the square of the first moment up to $o(1)$, proving the last relation in (\ref{N(a).1}).


$\square$

\medskip

\begin{proof}[Proof of Theorem \ref{t:max}]
Let $\varphi\in C_c^+(\mathbb R\times(-A,\infty))$. To prove the theorem, it is enough to show convergence of
the Laplace functional
\begin{equation}\label{main_max}
 I_\varphi=\mathbb E_{\rm KR}\exp\left\{-\sum_j
 \varphi(v_j,K_Np_j-\alpha_N)\right\}\longrightarrow
 \exp\left\{-\int_{\mathbb R}\frac{e^{-v^2}}{\sqrt\pi}\,dv
 \int_{\mathbb R}e^{-x}\bigl(1-e^{-\varphi(v,x)}\bigr)\,dx\right\}.
\end{equation}
Set
\[
 \mathcal Q_\varphi(r,\zeta):=\prod_{j=1}^N\left(c_Na_j(r,\zeta)
 \int_0^\infty e^{-c_Na_j(r,\zeta)p-\varphi(v_j,K_Np-\alpha_N)}\,dp\right),
\]
where $a_j(r,\zeta)=(\tilde b_j-r/2)^2+\zeta$. With
\[
 q_j(r,\zeta):=\frac{a_j(r,\zeta)}{\eta_N^2}e^{-\alpha_Na_j(r,\zeta)/\eta_N^2}
 \int_{-\alpha_N}^\infty e^{-a_j(r,\zeta)x/\eta_N^2}
 \bigl(1-e^{-\varphi(v_j,x)}\bigr)\,dx,
\]
the change of variables $x=K_Np-\alpha_N$ and $c_N/K_N=\eta_N^{-2}$ give
\begin{equation}\label{Q_cal}
 \mathcal Q_\varphi(r,\zeta)=\prod_{j=1}^N(1-q_j(r,\zeta)).
\end{equation}
Doing the same transformations with integrals
\[
c_Na_j(r,\zeta)
 \int_0^\infty p e^{-c_Na_j(r,\zeta)p-\varphi(v_j,K_Np-\alpha_N)}\,dp,
\]
we get that integrating with respect to $p_i$ and  arguing as in the proof of Theorem \ref{t:IPR}, we can bound the diagonal contribution by $o(|g_*|^2)$,
and write
\begin{align*}
&I_\varphi=\frac{\langle \mathcal Q_\varphi(r,\zeta) \hat g(\zeta,r,y_1) \hat g(\zeta,r,y_2)\rangle}{\langle g(\zeta,r,y_1)g(\zeta,r,y_2)\rangle}+o(1)\\
&\hat g(\zeta,r,y)=c_N^{-1}\sum_j\frac{a_j^{-1}}{(\tilde b_j+t_N+iy)}
\frac{1-a_j^2\eta_N^{-4}\int_{-\alpha_N}^\infty(x+\alpha_N)e^{-a_j\eta_N^{-2}(x+\alpha_N)}(1-e^{-\varphi(v_j,x)} )dx}
{1-a_j\eta_N^{-2}\int_{-\alpha_N}^\infty e^{-a_j\eta_N^{-2}(x+\alpha_N)}(1-e^{-\varphi(v_j,x)} )dx}
\end{align*}
Choose $R,T$ so that $\operatorname{supp}\varphi\subset[-R,R]\times[-T,T]$.

 Using (\ref{LLN3}) and (\ref{a_n_eq}), we get
\begin{align}\notag
&|g-\hat g|\le CN^{-1}\eta_N^{-2}\sum_j\int_{-T}^{T}dx e^{-\alpha_N}(\alpha_N+x) 
e^{-\eta_N^{-2}(\alpha_N+x)(\tilde b_j-r/2)^2}\\ \notag
\le &C\eta_N^{-2}\alpha_N\int_{-T}^{T}dx  N^{-1}e^{-\alpha_N}\sum_j e^{-\eta_N^{-2}(\alpha_N+x)(\tilde b_j-r/2)^2}\\ \notag
\le &C\rho(E)\frac{\alpha_N}{\eta_N\sqrt{\alpha_N}}e^{-\alpha_N}\le C\alpha_N(\eta_NK_N)^{-1}\le C\alpha_N(N\eta_N^2)^{-1}\le C|g_*|\alpha_N(N\eta_N)^{-1}
=|g_*|o(1)\\ \label{b_g_h}
&\Rightarrow \hat g(\zeta,r,y)=g(\zeta,r,y)(1+o(1))=g_*(1+o(1)).
\end{align}
Observe, that uniformly in $U_N$, $q_j=0$ unless $|v_j|\le R$, and on these indices
\[
 \frac{a_j(r,\zeta)}{\eta_N^2}=1+\frac{v_j^2}{\alpha_N}+o(\alpha_N^{-1}).
\]
Therefore,
\begin{align}\label{qsum}
 \sum_jq_j(r,\zeta)
 =e^{-\alpha_N}\sum_{|v_j|\le R}e^{-v_j^2}
 \int_{\mathbb R}e^{-x}\bigl(1-e^{-\varphi(v_j,x)}\bigr)\,dx\,[1+o(1)].
\end{align}
Notice that (\ref{LLN4}) with expansion of $\rho(b)$ at $b=E-t_N+v\eta_N/\sqrt{\alpha_N}$ give for general $F$ satisfying conditions of (\ref{LLN4})
\[
 \frac{\sqrt{\alpha_N}}{N\eta_N}\sum_jF(v_j)
 =\rho(E)\int_{\mathbb R}F(v)\,dv+o(1).
\]
Although the function
\[
F(v):=e^{-v^2}\int_{\mathbb R}e^{-x}\bigl(1-e^{-\varphi(v,x)}\bigr)\,dx
\]
need not be $C^1$, it is continuous and compactly supported. Hence, we can uniformly approximate it by functions in $C_c^1(\mathbb R)$. The error in the corresponding empirical average is negligible, uniformly on \(\Omega_N^{(2)}\), which can be obtained by the standard argument with application of (\ref{LLN4}) to the fixed non-negative cutoff 
$\chi\in C_c^1(\mathbb R)$ with $\chi\ge 1$ on a fixed compact set containing the supports of $F$ and all approximants.
Therefore, (\ref{LLN4}) applies to such $F$ as well.

Using this in \eqref{qsum}, together with \eqref{a_n_eq} and
$\eta_N=\pi\rho(E)\tau_N^2[1+o(1)]$, gives
\begin{align}\label{conv_q}
 \sum_jq_j(r,\zeta)\longrightarrow
 \int_{\mathbb R}\frac{e^{-v^2}}{\sqrt\pi}\,dv
 \int_{\mathbb R}e^{-x}\bigl(1-e^{-\varphi(v,x)}\bigr)\,dx .
\end{align}
Moreover $\max_j|q_j|=O(e^{-\alpha_N})=o(1)$, hence
$\sum_j|q_j|^2=o(1)$. Thus taking $\log$ of (\ref{Q_cal}) and expanding it up to the second term gives, uniformly in $U_N$,
\begin{equation}\label{Q-lim}
 \mathcal Q_\varphi(r,\zeta)=
 \exp\left\{-\int_{\mathbb R}\frac{e^{-v^2}}{\sqrt\pi}\,dv
 \int_{\mathbb R}e^{-x}\bigl(1-e^{-\varphi(v,x)}\bigr)\,dx\right\}(1+o(1)).
\end{equation}
Now (\ref{main_max}) can be obtained using (\ref{Q-lim}), (\ref{b_g_h}), and Lemma \ref{l:1}.
\end{proof}

\begin{proof}[Proof of Corollary \ref{c:max}.] Define 
\begin{align}\label{M_N}
\mathcal M_N(t)=&\sum_{j}\theta( K_Np_{j}-\alpha_N-t)=\sum_j\theta_j
\end{align}
Fix any $x$ and study
\begin{align*}
F_t(x)=\mathbb{E}_{\text{KR}}[e^{-x\mathcal M_N(t)}].
\end{align*}
To show $K_Np_{\max}-\alpha_N \Longrightarrow \mathcal{G}$, it is enough to prove that for any fixed $t, x$
\begin{equation}\label{F_t_conv}
F_t(x)\to e^{-(1-e^{-x})e^{-t}}.
\end{equation}
Integrating with respect to $\bar p$ and using the same argument as in Theorem \ref{t:IPR} to treat the diagonal terms, we obtain that
\begin{align*}
F_t(x)=\frac{\langle g(\zeta,r,y_1)g(\zeta,r,y_2)\prod (1-(1-e^{-x})e^{-\eta_N^{-2}(\alpha_N+t)((\tilde b_j-r/2)^2+\zeta)})\rangle}{\langle g(\zeta,r,y_1)g(\zeta,r,y_2)\rangle}(1+o(1)).
\end{align*}
Since
\begin{align*}
\Sigma_1=&\sum_j \log\Big(1-(1-e^{-x})e^{-\eta_N^{-2}(\alpha_N+t)((\tilde b_j-r/2)^2+\zeta)}\Big)\\
=&-(1-e^{-x})\sum_j e^{-\eta_N^{-2}(\alpha_N+t)((\tilde b_j-r/2)^2+\zeta)}+\tilde R(\zeta,r),\\
|\tilde R|\le & C e^{-\alpha_N}\sum_j \Big|e^{-\eta_N^{-2}(\alpha_N+t)((\tilde b_j-r/2)^2+\zeta)}\Big|,
\end{align*}
using (\ref{LLN3}), we obtain uniformly in $U_N$ w.h.p.
\begin{align*}
\Sigma_1=-Ne^{-\alpha_N-t}(1-e^{-x})\int \rho(b) e^{-\eta_N^{-2}(\alpha_N+t)(b-E-r/2)^2}db (1+o(1)).
\end{align*}
Note that 
\[
\eta_N^{-2}(\alpha_N+t)\Im\zeta
=O\!\left(\frac{\log^2N}{\sqrt{N\tau_N^2}}\right)=o(1)
\qquad\text{in }U_N
\]
so the estimate (\ref{LLN3}) applies.

Doing the change $u=(b-E-r/2)\sqrt{\alpha_N+t}/\eta_N$ and expanding $\rho(b)=\rho(E+r/2+u\eta_N/\sqrt{\alpha_N+t} )$ around $E$ we get 
\begin{align*}
\frac{\sqrt{\alpha_N+t}}{\eta_N}\int \rho(b) e^{-\eta_N^{-2}(\alpha_N+t)(b-E-r/2)^2} db=\sqrt{\pi}\rho(E)+o(1).
\end{align*}
Because of Lemma \ref{l:sol}, \eqref{K_N}, and \eqref{a_n_eq}, we have
\[
\dfrac{Ne^{-\alpha_N} \sqrt{\pi}\rho(E)\eta_N}{\sqrt{\alpha_N+t}}=1+o(1),
\]
thus
\begin{align*}
\prod (1-(1-e^{-x})e^{-\eta_N^{-2}(\alpha_N+t)((\tilde b_j-r/2)^2+\zeta)})=e^{-(1-e^{-x})e^{-t}}(1+o(1)),
\end{align*}
which gives (\ref{F_t_conv}).

\end{proof}

\medskip
\begin{proof}[Proof of Theorem \ref{t:min}]
Fix $f\in C_c^+((0,\infty)\times\mathbb R)$ and put
\[
 H_f(b):=\int_0^\infty(1-e^{-f(x,b)})\,dx.
\]
To prove the theorem, it is enough to show that
\begin{align}\label{main_min}
 \mathbb E_{KR}\big[e^{-\sum\limits_i f(\mathcal A_N p_i,b_i)}\big]
 \longrightarrow
 \exp\left\{-\int_{\mathbb R} H_f(b)
 d\nu_E(b)\right\}.
\end{align}
Recall the notation $a_j(r,\zeta)=(\tilde b_j-r/2)^2+\zeta$.
Define 
\begin{align}\label{P}
 \mathcal P_f(r,\zeta):=
 \prod_{j=1}^N\Big(
 c_Na_j(r,\zeta) \int_0^\infty e^{-c_Na_j(r,\zeta)p_j- f(\mathcal A_N p_j,b_j)}dp_j\Big):=\prod_{j=1}^NP_{f,j}(r,\zeta)
\end{align}

Notice
\begin{align*}
P_{f,j}(r,\zeta)=c_Na_j(r,\zeta) \int_0^\infty e^{-c_Na_j(r,\zeta)p_j- f(\mathcal A_N p_j,b_j)}dp_j=\frac{c_Na_j(r,\zeta)}{\mathcal A_N} \int_0^\infty e^{-c_Na_j(r,\zeta)x/\mathcal A_N- f(x,b_j)}dx.
\end{align*}
Taking
\[
 \lambda_j(r,\zeta):=\frac{c_Na_j(r,\zeta)}{\mathcal A_N},
\]
we can rewrite, omitting arguments $(r,\zeta)$ for simplicity,  the previous integral as
\begin{align}\notag
P_{f,j}(r,\zeta)=&\lambda_j\int_0^\infty e^{-\lambda_jx- f(x,b_j)}dx
=\int_0^\infty\lambda_j e^{-\lambda_jx}\,dx
-\lambda_j\int_0^\infty
e^{-\lambda_jx}\bigl(1-e^{-f(x,b_j)}\bigr)\,dx\\ \label{P_f,j}
&=1-\lambda_j\int_0^\infty
e^{-\lambda_jx}\bigl(1-e^{-f(x,b_j)}\bigr)\,dx
\end{align}
Doing the same manipulations with 
\begin{align*}
c_Na_j(r,\zeta) \int_0^\infty p_je^{-c_Na_j(r,\zeta)p_j- f(\mathcal A_N p_j,b_j)}dp_j,
\end{align*}
we again get that integrating with respect to $p_i$ and using the same arguments as in Theorems \ref{t:IPR}, \ref{t:max} to show that the diagonal terms contribute only $o(|g_*|^2)$, 
\begin{align*}
&\mathbb E_{KR}\big[e^{-\sum\limits_i f(\mathcal A_N p_i,b_i)}\big]=
\frac{\langle \mathcal P_f(r,\zeta) \tilde g(\zeta,r,y_1)\tilde g(\zeta,r,y_2)\rangle}{\langle g(\zeta,r,y_1)g(\zeta,r,y_2)\rangle}+o(1),\\
&\tilde g(\zeta,r,y)=c_N^{-1}\sum\frac{a_j^{-1}}{\tilde b_{j}+t_N+iy}\frac{1-\lambda_j^2\int x e^{-\lambda_jx}(1-e^{-f(x,b_j)})dx}
{1-\lambda_j\int  e^{-\lambda_jx}(1-e^{-f(x,b_j)})dx}
%
\end{align*}
At the  saddle
\[
 \lambda_j(r_*,\zeta_*)=
\frac{\beta_j^2+\eta_N^2}{N\mu_2(E)}
 =:\lambda_{j,*},
 \qquad \sum_j\lambda_{j,*}=1+o(1).
\]
Indeed, on  $\Omega_N^{(2)}$  we have 
uniformly for $B_N\in\Omega_N$ and $E\in I$
\[
 \frac1N\sum_i\bigl(\beta_i^2+\eta_N^2\bigr)
 =\frac1N\sum_i(b_i-E)^2+o(1)=\mu_2(E)+o(1),
\]
In addition,
\begin{equation}\label{kappa}
\max_j \lambda_{j,*} =o(1),
\end{equation}
since 
\[
 \max_i(\beta_i^2+\eta_N^2)
 \leq C\max_i(b_i-E)^2+o(1)=o(N).
\]
Now notice
\[
\sum_j  \left|a_j(r,\zeta)-(\beta_j^2+\eta_N^2)\right|
 \leq |r-r_*|\sum_j|\beta_j|+
 \frac N4|r-r_*|^2+N|\zeta-\eta_N^2|,
\]
so uniformly in $U_N$
\begin{equation}\label{l_j_b}
 \sum_j|\lambda_j-\lambda_{j,*}|=o(1),
 \qquad \max_j|\lambda_j|=o(1).
\end{equation}
If $\mathrm{supp}_x\,f(x,b)\in [0,T]$, then
\begin{align*}
\int_0^\infty e^{-\lambda_jx}(1-e^{-f(x,b_j)})\,dx=O_T(1)\\
\int_0^\infty xe^{-\lambda_jx}(1-e^{-f(x,b_j)})\,dx=O_T(1),
\end{align*}
so
\[
\left|
\frac{1-\lambda_j^2\int xe^{-\lambda_jx}(1-e^{-f})\,dx}
{1-\lambda_j\int e^{-\lambda_jx}(1-e^{-f})\,dx}
-1\right|
\le C_T|\lambda_j|,
\]
which, using (\ref{l_j_b}),   gives 
\begin{equation}\label{b_g_tild}
|\widetilde g-g|
 \le C_Tc_N^{-1}\sum_j
 \left|\frac{a_j^{-1}}{\beta_j+iy}\right||\lambda_j|
 =o(|g_*|).
\end{equation}


On the other hand, since 
$\mathrm{supp}_x\,f(x,b)\in [0,T]$ 
 each factor in (\ref{P}) satisfies, uniformly in $j$,
\begin{equation}\label{P_fj_asymp}
P_{f,j}(\zeta,r)=1-\lambda_j\int_0^\infty
e^{-\lambda_jx}\bigl(1-e^{-f(x,b_j)}\bigr)\,dx
 =1-\lambda_jH_f(b_j)+O_T(|\lambda_j|^2).
\end{equation}
Taking logarithms and using that in $\Omega^{(2)}_N$ by (\ref{mu_2.1})
\begin{align*}
\sum_j|\lambda_j|^2<C\delta_N,
\end{align*}
we obtain
\begin{equation}\label{P_lim}
 \log\mathcal P_f(r,\zeta)
 =-\sum_j\lambda_{j}H_f(b_j)+o(1)
 \qquad\hbox{uniformly in }U_N.
\end{equation}
By the definition of $\Omega_N^{(2)}$ 
\begin{align*}
 \sum_j\lambda_{j,*}H_f(b_j)
 &=(N\mu_2)^{-1}\sum_j(\beta_j^2+\eta_N^2)H_f(b_j) \\
 &=\frac1{\mu_2(E)}\int_{\mathbb R}(b-E)^2H_f(b)\rho(b)\,db+o(1)=\int_{\mathbb R} H_f(b)
 d\nu_E(b)+o(1)
\end{align*}
Notice that although $H_f$ is not necessarily Lipschitz, we can approximate $H_f$ uniformly on its compact support by bounded Lipschitz functions, and use the total-mass bound from $\Omega_N^{(2)}$ for the approximation error.

Thus $$\mathcal P_f=e^{-\int_{\mathbb R} H_f(b)
 d\nu_E(b)}(1+o(1))$$ uniformly on $U_N$.

%
Now (\ref{main_min}) can be obtained using (\ref{P_lim}), (\ref{b_g_tild}), and Lemma \ref{l:1}.

\end{proof}

\begin{proof}[Proof of Corollary~\ref{cor:min}]
For $t\geq0$, we repeat the preceding calculation from the proof of Theorem \ref{t:min} with 
an observable $\prod_j\theta{\{\mathcal A_Np_j-t\}}$. Instead of $\mathcal P_f$
it will give
\[
 \exp\Big\{-t\sum_j\lambda_j(r,\zeta)\Big\}.
\]
 But by (\ref{mu_2.1}) in $\Omega^{(2)}_N$ we have
\begin{equation*}
 \sum_j\lambda_j(r,\zeta)=1+o(1)
 \qquad\hbox{uniformly in }U_N,
\end{equation*}
and the analogue of estimate (\ref{b_g_tild}) on a new $\tilde g$  can be obtained similarly,
so
\[
 \mathbb P_{\rm KR}\{\mathcal A_Np_{\min}>t\}\longrightarrow e^{-t}.
\]
\end{proof}

\section{Discussion and outlook}
\label{sec:discussion}

The results give a law-level description of the principal one-eigenvector
occupation observables in the non-ergodic delocalized RP phase.  The
self-consistent Lorentzian weights identify an effective resonant set of size
$K_N(E)\asymp N^{2-\gamma}$, while the local exponential factors determine
the fluctuations about that profile.  Their combination explains both the
transition of the inverse participation sums at $q=1/2$ and the two linear
pieces of the amplitude-counting spectrum.

The two extreme processes reveal a pronounced asymmetry.  Large occupations
are selected from the resonant window $|b_i-E|=O(\eta_N)$ and exhibit the
Poisson/Gumbel statistics of many nearly independent exponential candidates,
modified by the Lorentzian envelope.  By contrast, the smallest occupations
are generated by the small-value tail over the full collection of sites; the
limiting diagonal mark is biased by $(b-E)^2$.  In particular, the maximum
and minimum do not arise from two tails of a single homogeneous sample.

The absence-of-freezing corollary makes the relation to REM and logREM
phenomenology precise at fixed $q$.  The upper extreme process is REM-like,
but no fixed positive participation sum condenses onto the largest
coordinate.  Questions in which $q$ itself grows with $N$, and possible
Poisson--Dirichlet limits for auxiliary high-$q$ escort measures, require a
different scaling and are not asserted here.

This paper treats one eigenvector at one regular bulk energy.  Natural next
problems include joint occupation laws for two nearby eigenvectors of the same
matrix, edge and crossover regimes, and the corresponding statements in the
orthogonal and symplectic symmetry classes.  The exact projective
Kac--Rice representation developed here provides a possible starting point
for these extensions.

\section*{Acknowledgements}

The research of YVF was supported by UKRI grant UKRI1015,
``Non-Hermitian Random Matrices: Theory and Applications.''    MS completed the work during her stay as Royal Society Wolfson Visiting Fellow (RSWVF/25/R2/1008) at the School of Mathematics at the University of Bristol. TS was supported in part by Alfred P. Sloan Foundation grant FG-2022-18916 and the National Science Foundation grant DMS-2346379.

The authors acknowledge assistance from ChatGPT (OpenAI) for performing consistency checks of the proofs, for critical reading and improving the clarity of the presentation, and designing numerical checks.  All scientific arguments, results, and conclusions
remain the responsibility of the authors.

\appendix
\section{Appendix}

Let $I$ be a compact interval on which $0<c_I\leq\rho\leq C_I$, and assume
that $\rho\in C^3$ in a fixed $\delta$-neighbourhood of $I$, with sufficient decay to
make its Hilbert transform well defined.  Put
\begin{align}\label{epsilon_N}
 \eta_*:=\inf_{E\in I}(\pi\tau_N^2\rho(E))\asymp\tau_N^2,
 \qquad \varepsilon_N:=\frac{\log N}{\sqrt{N\eta_*}}.
\end{align}
Note that $\varepsilon_N=o(1)$ because
of \eqref{tau}.

\begin{lemma}
\label{l:int}
For every fixed compact interval $J\subset\mathbb R$, uniformly for
$a\in J$ and as $\eta\downarrow0$
\begin{align}
 \int\frac{\rho(x)\,dx}{(a-x)^2+\eta^2}
 &=\frac{\pi\rho(a)}{\eta}+O(1),\label{of:L1}\\
 \int\frac{\rho(x)\,dx}{((a-x)^2+\eta^2)^2}
 &=\frac{\pi\rho(a)}{2\eta^3}+O(\eta^{-1}),\label{of:L2}\\
 \int\frac{(a-x)\rho(x)\,dx}{(a-x)^2+\eta^2}
 &=H_\rho(a)+O(\eta),\label{of:H1}\\
 \Big|\int\frac{(a-x)\rho(x)\,dx}{((a-x)^2+\eta^2)^2}\Big|
 &\le C/\eta,\label{of:H2}
\end{align}
where $H_\rho(a)=\operatorname{p.v.}\int\rho(x)/(a-x)\,dx$.  Consequently,
\begin{equation}\label{of:L2-scaled}
 \int\frac{\eta^2\rho(x)\,dx}{((a-x)^2+\eta^2)^2}
 =\frac{\pi\rho(a)}{2\eta}+O(\eta).
\end{equation}
\end{lemma}

\begin{proof}
Fix $J$.  Choose $\delta>0$ so that the closed $2\delta$-neighbourhood of $J$ lies in
the region where $\rho$ is $C^3$. Write $u=x-a$. Clearly, $|u|>\delta$ contributes $O(1)$ to the integrals in \eqref{of:L1} -- \eqref{of:L2} and (\ref{of:H2}), so we need to consider $|u|\le \delta$. Using $\rho\in C^3$, we can expand
\[
 \rho(a+u)=\rho(a)+u\rho'(a)+u^2R_a(u),
 \qquad \sup_{a,\,|u|\leq \delta}|R_a(u)|<\infty.
\]
For the first kernel, the constant term in this Taylor expansion gives
\[
 \rho(a)\int_{-\delta}^{\delta}\frac{du}{u^2+\eta^2}
 =\frac{\pi\rho(a)}{\eta}+O(1).
\]
The linear term is an odd function, thus vanishes after the integration, and the remainder contributes at most
\[
 C\int_{-\delta}^{\delta}\frac{u^2\,du}{u^2+\eta^2}=O(1).
\]
This proves \eqref{of:L1}.  Similarly,
\[
 \rho(a)\int_{-\delta}^{\delta}\frac{du}{(u^2+\eta^2)^2}
 =\frac{\pi\rho(a)}{2\eta^3}+O(1).
\]
Again the linear term cancels, while
\[
 \int_{-\delta}^{\delta}\frac{u^2\,du}{(u^2+\eta^2)^2}=O(\eta^{-1}).
\]
 This proves \eqref{of:L2}, and thus (\ref{of:L2-scaled}).
 
 Using $\rho(a+u)=\rho(a)+uR(u)$ with $\sup_{a,\,|u|\leq \delta}|R(u)|<\infty$, one can similarly obtain (\ref{of:H2}).

To prove (\ref{of:H1}), write
\[
 \frac{a-x}{(a-x)^2+\eta^2}-\frac1{a-x}
 =-\frac{\eta^2}{(a-x)((a-x)^2+\eta^2)}.
\]
For $|u|\le \delta$, we can again use the Taylor expansion.
The first term is vanishing, and the remaining integrand
is bounded by $C\frac{\eta^2}{u^2+\eta^2}$,
whose integral is $O(\eta)$.  The region $|u|> \delta$ contributes $O(\eta^2)$, so we get \eqref{of:H1}.

\end{proof}

\begin{proposition}
\label{p:LLN}
Fix a compact interval  $J\subset\mathbb R$, and choose $\varepsilon_N$ as in (\ref{epsilon_N}). For every $D>0$,  with probability at least
$1-N^{-D}$ the following hold: 

\begin{itemize}
\item  uniformly in $E\in I$, $r\in J$, $c\eta_*\leq\eta\leq C\eta_*$ for $s\in \{1,2\}$:
\begin{align}\label{LLN1}
 \frac1N\sum_i\frac{\tilde b_i-r/2}{((\tilde b_i-r/2)^2+\eta^2)^s}
 &=\int\frac{(b-E-r/2)\rho(b)\,db}{((b-E-r/2)^2+\eta^2)^s}
   +O(\eta^{-2s+2}\varepsilon_N);
   \end{align}
\item uniformly in
$E\in I$, $|r|\le R$, and $\zeta\in \eta^2+i\mathbb R$, $|\Im \zeta|\le C$, $c\eta_*\leq\eta\leq C\eta_*$, for $s\in \{1/2,1,2\}$
\begin{equation}\label{LLN2}
N^{-1}\sum_j\frac{1}{((\tilde b_{j}-r/2)^2+\zeta)^s}=\int \frac{\rho(b)db}{((b-E-r/2)^2+\zeta)^s}+O(\eta^{-2s+1}\varepsilon_N);
\end{equation}   
   
\item  If $\{\nu_N\}$ is a sequence that grows more slowly than any positive
power of $N$, then, for every fixed $0<\delta<2$, uniformly in
$E\in I$ and $|r|\le R\tau_N^2$, we have
\begin{align}\label{LLN3}
\sqrt{\nu_N}N^{-1+\delta/2}\sum_je^{-N^{\delta}\nu_N(\tilde b_{j}-r/2)^2}=&
\sqrt{\nu_N}N^{\delta/2}\int db\rho(b)e^{-N^{\delta}\nu_N(b-E-r/2)^2}\\
&+O(\nu_N^{1/4}N^{-(1-\delta/2)/2}\log N).
\notag\end{align}
Moreover, for every fixed $C_0>0$, the estimate holds uniformly over
all choices of $\nu_N$ satisfying
\[
 1\leq \nu_N\leq(\log N)^{C_0}.
\]

In general, let $F\in C^1(\mathbb R)$ and suppose that there are constants
$C_F,R_F,\kappa>0$ such that
\begin{equation}\label{tail}
 |F(u)|+|F'(u)|\leq C_F(1+|u|)^{-1-\kappa},\qquad |u|>R_F.
\end{equation}
Then uniformly for
$E\in I$, $|t|\leq T\tau_N^2$, and
$c\eta_*\leq\eta\leq C\eta_*$,
\begin{equation}\label{LLN4}
 \frac{1}{N}\sum_{i=1}^NF\left(\sqrt{\nu_N}\frac{b_i-E+t}{\eta}\right)
 =\int_{\mathbb R}F\left(\sqrt{\nu_N} \frac{b-E+t}{\eta}\right)\rho(b)\,db
 +o(\eta/\sqrt{\nu_N}).
\end{equation}

\item uniformly in
$E\in I$ and $r,\zeta,y_1,y_2\in U_N$
\begin{align}\label{LLN5}
 N^{-1}\sum_{j}e^{-(\tilde b_{j}-r/2)^{2}}=\int \rho(b)e^{-(b-E)^{2}}db+O(|r|)+O(N^{-1/2}\log N);
\end{align}
\end{itemize}

\end{proposition}

\begin{proof}
For a bounded measurable function $f$, Bernstein's inequality in the form
used here is
\begin{equation}\label{of:Bernstein-empirical}
 \mathbb P\Big[\Big|\frac1N \sum_{i=1}^N
 f(b_i)-\mathbb Ef(b_1)\Big|>u\Big]
 \leq2\exp\Big\{-\frac{Nu^2}
 {2(\mathbb E|f(b_1)|^2+ u\|f\|_\infty/3)}\Big\}.
\end{equation}
We apply this inequality at a fixed pair $(r,\eta)$.
For
\[
 f_s(b)=\frac{b-E-r/2}{((b-E-r/2)^2+\eta^2)^s},\quad s=1,2,
\]
one has, uniformly in $r\in J$, $c\eta_*\leq\eta\leq C\eta_*$,
\[
 \|f_s\|_\infty\leq\eta^{1-2s},
 \qquad
 |\mathbb E[f_s(b)^2]|
 \leq C\int_{\mathbb R}\frac{(b-E-r/2)^2 db}{((b-E-r/2)^2+\eta^2)^{2s}}
 \leq C\eta^{3-4s}.
\]
Thus, taking $u_s=\eta^{2-2s}\varepsilon_N$, we get 
\[
 \mathbb P\Big[\Big|\frac1N\sum_i f_s(b_i)-\mathbb Ef_s(b_1)\Big|
 >u_s\Big]
 \leq2\exp\{-cN\eta_*\varepsilon_N^2\}
 =2e^{-C\log^2N}.
\]
This implies (\ref{LLN1}) for fixed $(r,\eta)$. 

It remains to make the estimates uniform. To this end, we choose a rectangular mesh in
$r\in J$, $c\eta_*\leq\eta\leq C\eta_*$ of spacing $h=e^{-C\log^{3/2}N}$. It has at most $e^{C\log^{3/2}N}$ points, and since the derivative of the function 
$f_s$ with respect to $(r,\eta)$ grow at most polynomially in $N$, we obtain (\ref{LLN1}).

The bound (\ref{LLN2}) can be proved similarly by considering real and imaginary part of
\[
 g_s(b)=\frac{1}{((b-E-r/2)^2+\zeta)^s}.
\]
For $s=1/2$, the corresponding summand $g$ satisfies
$\|g\|_\infty\leq\eta^{-1}$ and
$\mathbb E|g(b_1)|^2\leq C\log(1/\eta)$; Bernstein's inequality
with $u=\varepsilon_N$ gives the stated estimate.

To prove  (\ref{LLN3}) we consider
\begin{align*}
&h_t(b)=\sqrt{\nu_N}N^{\delta/2}e^{-N^\delta\nu_N(b-E-r/2)^2},
\end{align*}
with
\[
\|h_t\|_\infty\le \sqrt{\nu_N}N^{\delta/2} \quad  \mathbb E[|h_t(b)|^2]\le\sqrt{\nu_N}N^{\delta/2} 
\]
and $u=\nu_N^{1/4}N^{-(1-\delta/2)/2}\log N$. To obtain uniformity in 
$1\leq\nu\leq(\log N)^{C_0}$, we also apply the
fixed-parameter estimates on a polynomial mesh and use the polynomial
bounds on the parameter derivatives to control the interpolation
error. (\ref{LLN4}) can be proved similarly by applying Bernstein inequality on $|u|\le L$, and then controlling $|u|>L$ 
by dyadic shells, and let $L\to\infty$. 
\end{proof}

%

\begin{lemma}
\label{l:sol}
On the event of Proposition~\ref{p:LLN}, uniformly for
$E\in I$, the equations \eqref{of:pastur} have a unique solution $(t_N(E), \eta_N(E))$ in
$|t|\leq C\tau_N^2$, $c\tau_N^2\leq\eta\leq C\tau_N^2$, and
\begin{align}
 \eta_N(E)&=\pi\tau_N^2\rho(E)
 \bigl[1+O(\tau_N^2+\varepsilon_N)\bigr],\label{of:eta-asymp}\\
 t_N(E)&=\tau_N^2H_\rho(E)
 +O\bigl(\tau_N^2(\tau_N^2+\varepsilon_N)\bigr).
 \label{of:t-asymp}
\end{align}
where $\varepsilon_N$ was defined in (\ref{epsilon_N}).
\end{lemma}

\begin{proof}
For $a\in I+[-\delta,\delta]$ and $\eta>0$, introduce 
\[
 L_N(a,\eta)=\frac{\tau_N^2}{N}\sum_{i=1}^N
 \frac1{(a-b_i)^2+\eta^2},
 \qquad
 M_N(a,\eta)=\frac{\tau_N^2}{N}\sum_{i=1}^N
 \frac{a-b_i}{(a-b_i)^2+\eta^2}.
\]
Then \eqref{of:pastur} is equivalent to
\begin{equation}\label{of:self-consistent-system}
 L_N(E-t,\eta)=1,
 \qquad t=M_N(E-t,\eta).
\end{equation}
Proposition~\ref{p:LLN}, followed by
Lemma~\ref{l:int}, gives uniformly in $|t|\leq C\tau_N^2$, $c\tau_N^2\leq\eta\leq C\tau_N^2$
\begin{align}
 L_N(a,\eta)
 &=\frac{\pi\tau_N^2\rho(a)}{\eta}
   +O(\tau_N^2+\varepsilon_N),\label{of:LN-expansion}\\
 M_N(a,\eta)
 &=\tau_N^2H_\rho(a)
   +O\bigl(\tau_N^2\eta+\tau_N^2\varepsilon_N\bigr).
 \label{of:MN-expansion}
\end{align}
Put
\[
 t=\tau_N^2u,\qquad \eta=\tau_N^2v,
\]
where $|u|\le C$ and $c\le v\le C$.  Define
\[
 \mathfrak F_N(u,v;E)=
 \begin{pmatrix}
 L_N(E-\tau_N^2u,\tau_N^2v)-1\\[2mm]
 \tau_N^{-2}\bigl[\tau_N^2u-
 M_N(E-\tau_N^2u,\tau_N^2v)\bigr]
 \end{pmatrix}.
\]
By \eqref{of:LN-expansion}--\eqref{of:MN-expansion},
$\mathfrak F_N$ is uniformly close with an error $\omega_N=\tau_N^2+\varepsilon_N=o(1)$
to
\[
 \mathfrak F_0(u,v;E)=
 \begin{pmatrix}
 \pi\rho(E)/v-1\\ u-H_\rho(E)
 \end{pmatrix}.
\]
The same conclusion holds for the first derivatives. Indeed, since
\[
\partial_a\frac1{x^2+\eta^2}
=-\frac{2x}{(x^2+\eta^2)^2},
\qquad
\partial_\eta\frac1{x^2+\eta^2}
=-\frac{2\eta}{(x^2+\eta^2)^2},
\]
and
\[
\partial_a\frac{x}{x^2+\eta^2}
=\frac{\eta^2-x^2}{(x^2+\eta^2)^2}
=\frac{2\eta^2}{(x^2+\eta^2)^2}
-\frac1{x^2+\eta^2},
\]
\[
\partial_\eta\frac{x}{x^2+\eta^2}
=-\frac{2\eta x}{(x^2+\eta^2)^2},
\]
one can apply Lemmas \ref{l:int} and Proposition \ref {p:LLN}.  
Hence
\[
 \|\mathfrak F_N-\mathfrak F_0\|_{C^1}=O(\tau_N^2+\varepsilon_N)=O(\omega_N)
\]
uniformly on the rectangle.

The unique zero of $\mathfrak F_0$ is
\[
 z_0(E):=(u_0(E),v_0(E))=(H_\rho(E),\pi\rho(E)).
\]
Put $A_E=D\mathfrak F_0(z_0(E);E)$.  The matrices $A_E$ are uniformly
invertible for $E\in I$.  Choose $\delta>0$ so small that the closed
$\delta$-ball around $z_0(E)$ lies in the interior of the displayed
rectangle for every $E\in I$ and
\[
 \sup_{\substack{E\in I\\|z-z_0(E)|\le\delta}}
 \|A_E^{-1}(D\mathfrak F_0(z;E)-A_E)\|\le\frac14.
\]
By the $C^1$ estimate above, for all sufficiently large $N$,
\begin{equation}\label{contraction-estimate}
 \sup_{\substack{E\in I\\|z-z_0(E)|\le\delta}}
 \|A_E^{-1}(D\mathfrak F_N(z;E)-A_E)\|\le\frac12.
\end{equation}
By compactness and the fact that $\mathfrak F_0$ has no other zero in the
displayed rectangle,
\[
 c_\delta:=\inf_{\substack{E\in I,\ (u,v)\ \text{in the displayed rectangle}\\
 |(u,v)-(u_0(E),v_0(E))|\geq\delta}}
 |\mathfrak F_0(u,v;E)|>0.
\]
The uniform $C^0$ estimate above implies, for all sufficiently large $N$,
that $\mathfrak F_N$ has no zero outside this $\delta$-ball.

For completeness, define
\[
 T_{N,E}(z):=z-A_E^{-1}\mathfrak F_N(z;E).
\]
By \eqref{contraction-estimate}, $T_{N,E}$ is a contraction on
$B_\delta(z_0(E))$.  Moreover,
\[
 |T_{N,E}(z_0(E))-z_0(E)|
 \le \|A_E^{-1}\|\,|\mathfrak F_N(z_0(E);E)|
 \le C\omega_N.
\]
Hence, for a sufficiently large fixed $C$, $T_{N,E}$ maps
$B_{C\omega_N}(z_0(E))$ into itself.  The contraction mapping theorem gives a
zero of $\mathfrak F_N$ in this ball.  Finally, if $z$ is any zero in
$B_\delta(z_0(E))$, the mean-value formula and
\eqref{contraction-estimate} give
\[
 |z-z_0(E)|
 \le 2\|A_E^{-1}\|\,|\mathfrak F_N(z_0(E);E)|
 \le C\omega_N.
\]
Thus every zero in the $\delta$-ball belongs to the contraction ball, and
the zero is unique.
Thus the zero of \eqref{of:self-consistent-system} is unique in the whole
displayed rectangle and
\[
 u=H_\rho(E)+O(\omega_N),
 \qquad v=\pi\rho(E)[1+O(\omega_N)],
\]
uniformly in $E$.  This is precisely
\eqref{of:eta-asymp}--\eqref{of:t-asymp}.


\end{proof}

\end{document}